\documentclass[twocolumn,superscriptaddress,nobalancelastpage,amsmath,amssymb]{revtex4-1}
\usepackage{graphicx}
\graphicspath{{/}{Figures/}}
\usepackage{bm}

\usepackage{xcolor}

\newcommand {\ii}      {\mathrm{i}}
\newcommand {\ee}      {\mathrm{e}}
\newcommand {\nm}      {\,\mathrm{nm}}
\newcommand {\meV}     {\,\mathrm{meV}}

\renewcommand {\vec}   {\mathbf}

\newcommand {\ket}[1]      {\lvert#1\rangle}

\newcommand {\bramidket}[3] {\langle#1\vert#2\vert#3\rangle}

\newcommand {\avg}[1]   {\langle#1\rangle}

\newcommand {\eqn}     {Eq.~}

\newcommand {\kdotp}   {\vec{k}\cdot\vec{p}}
\newcommand {\muB}     {\mu_B}
\newcommand {\kB}      {k_B}

\begin{document}


\title{Emergent quantum Hall effects below $50$\,mT in a two-dimensional topological insulator}

\author{Saquib Shamim}
\altaffiliation{These authors contributed equally to this work}
\affiliation{Physikalisches Institut (EP3), Universit\"{a}t W\"{u}rzburg, Am Hubland, 97074 W\"{u}rzburg, Germany}
\affiliation{Institute for Topological Insulators, Am Hubland, 97074 W\"{u}rzburg, Germany}

\author{Wouter Beugeling}
\altaffiliation{These authors contributed equally to this work}
\affiliation{Physikalisches Institut (EP3), Universit\"{a}t W\"{u}rzburg, Am Hubland, 97074 W\"{u}rzburg, Germany}
\affiliation{Institute for Topological Insulators, Am Hubland, 97074 W\"{u}rzburg, Germany}

\author{Jan B\"{o}ttcher}
\altaffiliation{These authors contributed equally to this work}
\affiliation{Institut f\"ur Theoretische Physik und Astrophysik, Universit\"{a}t W\"{u}rzburg, Am Hubland, 97074 W\"{u}rzburg, Germany}

\author{Pragya Shekhar}
\affiliation{Physikalisches Institut (EP3), Universit\"{a}t W\"{u}rzburg, Am Hubland, 97074 W\"{u}rzburg, Germany}
\affiliation{Institute for Topological Insulators, Am Hubland, 97074 W\"{u}rzburg, Germany}

\author{Andreas Budewitz}
\affiliation{Physikalisches Institut (EP3), Universit\"{a}t W\"{u}rzburg, Am Hubland, 97074 W\"{u}rzburg, Germany}
\affiliation{Institute for Topological Insulators, Am Hubland, 97074 W\"{u}rzburg, Germany}

\author{Philipp Leubner}
\affiliation{Physikalisches Institut (EP3), Universit\"{a}t W\"{u}rzburg, Am Hubland, 97074 W\"{u}rzburg, Germany}
\affiliation{Institute for Topological Insulators, Am Hubland, 97074 W\"{u}rzburg, Germany}

\author{Lukas Lunczer}
\affiliation{Physikalisches Institut (EP3), Universit\"{a}t W\"{u}rzburg, Am Hubland, 97074 W\"{u}rzburg, Germany}
\affiliation{Institute for Topological Insulators, Am Hubland, 97074 W\"{u}rzburg, Germany}

\author{Ewelina M. Hankiewicz}
\affiliation{Institut f\"ur Theoretische Physik und Astrophysik, Universit\"{a}t W\"{u}rzburg, Am Hubland, 97074 W\"{u}rzburg, Germany}

\author{Hartmut Buhmann}
\affiliation{Physikalisches Institut (EP3), Universit\"{a}t W\"{u}rzburg, Am Hubland, 97074 W\"{u}rzburg, Germany}
\affiliation{Institute for Topological Insulators, Am Hubland, 97074 W\"{u}rzburg, Germany}

\author{Laurens W. Molenkamp}
\affiliation{Physikalisches Institut (EP3), Universit\"{a}t W\"{u}rzburg, Am Hubland, 97074 W\"{u}rzburg, Germany}
\affiliation{Institute for Topological Insulators, Am Hubland, 97074 W\"{u}rzburg, Germany}

\begin{abstract}

\end{abstract}

\maketitle

\textbf{The realization of the quantum spin Hall effect in HgTe quantum wells has led to the development of topological materials which, in combination with magnetism and superconductivity, are predicted to host chiral Majorana fermions. However, the large magnetization ($\sim$ a few tesla) in conventional quantum anomalous Hall system, makes it challenging to induce superconductivity. Here, we report two different emergent quantum Hall effects in HgTe quantum wells dilutely alloyed with Mn. Firstly, a novel quantum Hall state emerges from the quantum spin Hall state at an exceptionally low magnetic field of $\sim 50$\,mT. Secondly, tuning towards the bulk $p$-regime, we resolve multiple quantum Hall plateaus at fields as low as $20$--$30$\,mT, where transport is dominated by a van Hove singularity in the valence band. These emergent quantum Hall phenomena rely critically on the topological band structure of HgTe and their occurrence at very low fields make them an ideal candidate for interfacing with superconductors to realize chiral Majorana fermions.}

\section*{Introduction}

Though the quantum Hall (QH) effect has been understood as a consequence of topological order for nearly four decades~\cite{ThoulessEA1982,Wen1995}, the interest in the field of topology has rekindled in the last decade due to the successful realization of topological insulators (TIs)~\cite{KonigEA2007,HsiehEA2008,BruneEA2011}. These systems support gapless conducting states at the surface with an insulating bulk in the absence of an external magnetic field~\cite{QiZhang2011}. Among the known topological systems, HgTe is a versatile material as it can be tuned from a trivial insulator to a two- or three-dimensional (2D or 3D) TI by changing thickness and strain~\cite{KonigEA2007,BruneEA2011,LeubnerEA2016PRL}. Confinement in a HgTe quantum well (QW) above a critical thickness results in an inverted band structure and forms a 2D TI with quantum spin Hall (QSH) edge states protected by time-reversal symmetry~\cite{BernevigEA2006,KonigEA2007}. The band structure can be further engineered by alloying HgTe with magnetic or non-magnetic atoms. Non-magnetic dopants affect mainly the band gap while magnetic dopants can have more profound effects as they can break time-reversal symmetry. In case of HgTe QWs, a magnetically doped 2D TI can be realized by incorporating Mn atoms \cite{LiuEA2008PRL101}. In contrast to magnetically doped bismuth based TIs~\cite{ChangEA2013Science}, (Hg,Mn)Te is paramagnetic and hence, in the absence of an external magnetic field, the magnetization is zero~\cite{Furdyna1988,NovikEA2005}. A QSH state is therefore expected when the band structure is in the inverted regime. The interest in (Hg,Mn)Te arises from the prediction of a quantum anomalous Hall (QAH) state when Mn is magnetized~\cite{LiuEA2008PRL101}. It has been predicted that a chiral topological superconductor hosting chiral Majorana fermions can be realized when a chiral edge channel (either from a QH or a QAH state) is proximitized to a s-wave superconductor~\cite{QiEA2010PRB}. However, the large magnetic fields ($> 1$\,T) required for QH effects and the large magnetization ($\sim$ a few tesla) in conventional QAH systems (e.g., vanadium or chromium doped Bi-based topological systems) make it challenging to induce superconductivity and explore chiral Majorana fermions. It is thus significant to find systems where the QH effects can be observed at very low magnetic fields and low magnetization.

In this article, we show two emergent QH phenomena arising from different mechanisms at unusually low magnetic fields in (Hg,Mn)Te based 2D TIs. Firstly, when the chemical potential is tuned into the bulk gap (the `QSH regime') of a direct-gap 2D TI, we observe a $\nu=-1$ QH plateau ($\nu$ is the filling factor) which emerges from the QSH state in perpendicular magnetic fields as low as $50$\,mT. Band structure analysis demonstrates that this QH state forms due to splitting of the Dirac crossing and subsequent hybridization of one of the QSH edge states with the bulk states. The $\nu=-1$ plateau is located outside the topological gap (QSH regime), and is therefore fundamentally different from the QAH state predicted in Ref.~\cite{LiuEA2008PRL101}.

Secondly, when the chemical potential is tuned into the valence band ($p$-conducting) regime, we find that the transport is predominantly affected by the `camelback', i.e., the maximum in the valence band at a large momentum that arises as a consequence of hybridization between the subbands in an inverted band structure. At finite magnetic fields, the magnetization of Mn leads to corrections in the dispersion relation via the paramagnetic exchange interaction~\cite{NovikEA2005,LiuEA2008PRL101}. The interplay between this exchange interaction and the van Hove singularity (very large density of states (DOS)), induced by the camelback,  leads to rich LL structures: In direct-gap TIs, the large DOS at the camelback pins the chemical potential, which leads to formation of highly mobile carriers (mobility $\sim 10^6\,\mathrm{cm}^{2}\,\mathrm{V}^{-1}\,\mathrm{s}^{-1}$) at the $\Gamma$-point with an extremely low carrier density of $\sim 2\times 10^9\,\mathrm{cm}^{-2}$. This results in multiple resolved QH plateaus below $100$\,mT at low temperatures. Magneto-transport measurements in tilted magnetic fields support this pinning picture. In contrast, in indirect-gap TIs, the pinning mechanism permits QH plateaus in the bulk $p$-regime only at large magnetic fields.

The emergent QH phenomena rely critically on the preexistence of QSH states at zero field and the van Hove singularity in the DOS, both of which originate from the inverted band structure of HgTe. Since band inversion makes HgTe topological, our findings have a direct significant impact on the fundamental understanding of transport phenomena in all topological materials. Additionally, the occurrence of emergent QH effects in (Hg,Mn)Te QWs at extremely low fields makes this system an ideal platform for interfacing with s-wave superconductors to realize chiral Majorana fermions~\cite{QiEA2010PRB}.


\begin{figure}[tb]
\includegraphics[width=1\linewidth]{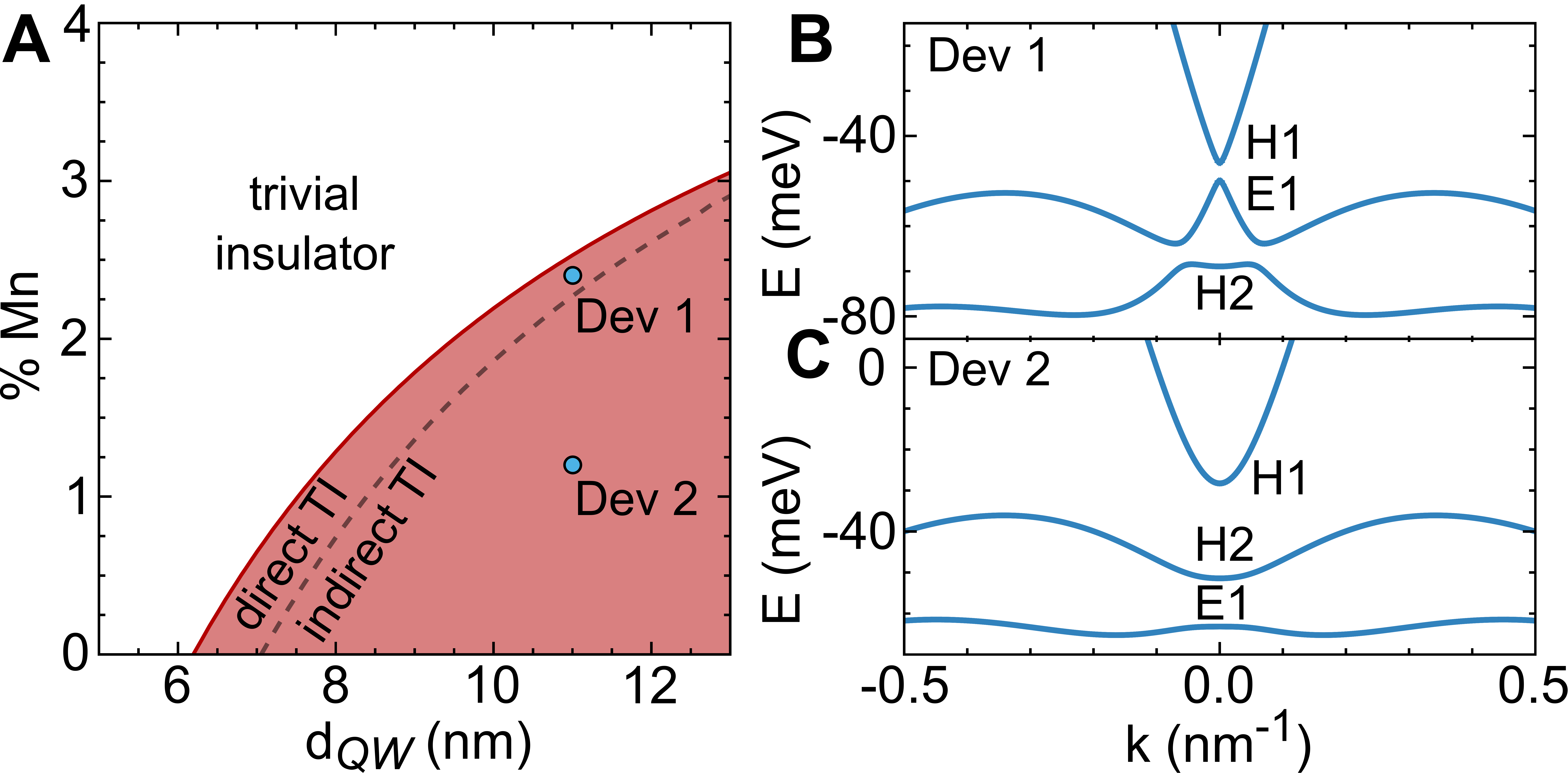}
\caption{\textbf{Band structure of (Hg,Mn)Te quantum wells.}
(\textbf{A}) Phase space obtained from $\kdotp$ calculations of (Hg,Mn)Te QWs of different thicknesses $d_{QW}$ and Mn concentrations. The solid red line indicates the transition from trivial (white) to topological (red) regime. Along this critical line, the E1 (electron-like) and the H1 (heavy-hole-like) subbands are energetically inverted.  The dotted line indicates the direct- to indirect-gap transition.
(\textbf{B}) The $\kdotp$ band structure of (Hg,Mn)Te QWs of thickness $11$\,nm with $2.4$\% Mn (Dev 1)  with a direct gap and
(\textbf{C}) $11$\,nm with $1.2$\% Mn (Dev 2) with an indirect gap, where H2 marks the second heavy-hole-like subband.
The energy $E$ is defined relative to the conduction band of unstrained bulk HgTe.
}
\label{FigPhaseSpace}
\end{figure}


\section{Results}

We explore the topological phase space of (Hg,Mn)Te QWs using $\mathbf{k}\cdot\mathbf{p}$ theory based on the $8\times8$-Kane Hamiltonian applied to bulk and strip geometries (with confinement in one and two directions, respectively), where the latter is necessary to describe both bulk and edge states. Despite this method being computationally heavy, we avoid using simplified band models such as the one proposed by Bernevig, Hughes, and Zhang (BHZ) \cite{BernevigEA2006}, since they do not capture the essential physics in the valence band, i.e., the camelback. The existence of the camelback in the band structure of these QWs has been known for many decades~\cite{PfeufferJeschke2000_thesis,OrtnerEA2002PRB} and has also attracted attention recently~\cite{LeubnerEA2016PRL,MinkovEA2016PRB}. The large DOS at the camelback has been shown to give rise to an interaction induced $0.5$ anomaly in the HgTe based topological quantum point contacts~\cite{StrunzEA2019}.

The transport behaviour of (Hg,Mn)Te QWs can be essentially understood from two properties of the zero-field dispersion: The band ordering at $k=0$, as well as the presence and positions of additional extrema in the dispersion at finite momenta. The first property is well-known for HgTe QWs, which exhibit a `trivial' band order below the critical thickness $d_\mathrm{c}\approx6.3$\,nm and host an inverted band structure above $d_c$, characterized by the QSH phase~\cite{BernevigEA2006}. Due to the influence of the Mn atoms on the band structure, this critical thickness varies with Mn concentration. In Fig.~1A, we map out the parameter space of band ordering in terms of QW thickness $d_\mathrm{QW}$ and Mn concentration: The white and red areas indicate the trivial and the inverted regime, respectively. Secondly, band structures can be distinguished by having either a direct or an indirect band gap depending on the position of the camelback. In Fig.~1A, the direct-indirect transition is represented by a dashed curve.

We focus on devices fabricated from two (Hg,Mn)Te QWs which are $11$\,nm thick with Mn concentration of $2.4$\% and $1.2$\% labelled as Dev $1$ and Dev $2$ in Fig.~1, respectively. The QWs are embedded between Hg$_{\mathrm{0.32}}$Cd$_{\mathrm{0.68}}$Te barriers on a lattice matched $\mathrm{Cd_{0.96}Zn_{0.04}Te}$ substrate and patterned into Hall bars of length $l=600\,\mathrm{\mu m}$ and width $w=200\,\mathrm{\mu m}$ (see Methods for fabrication details). The Mn atoms substitute Hg atoms isoelectrically ensuring that neither carrier doping nor degradation in mobility takes place. The $\mathbf{k}\cdot\mathbf{p}$ band structures show that Dev $1$ has a direct gap (Fig.~1B) while Dev $2$ has an indirect gap (Fig.~1C). We show results from Dev $1$ in the main text and from Dev $2$ in the Supplementary Information. All measurements are performed at temperature $T = 20$\,mK.


\begin{figure*}[tb]
\includegraphics[width=0.8\linewidth]{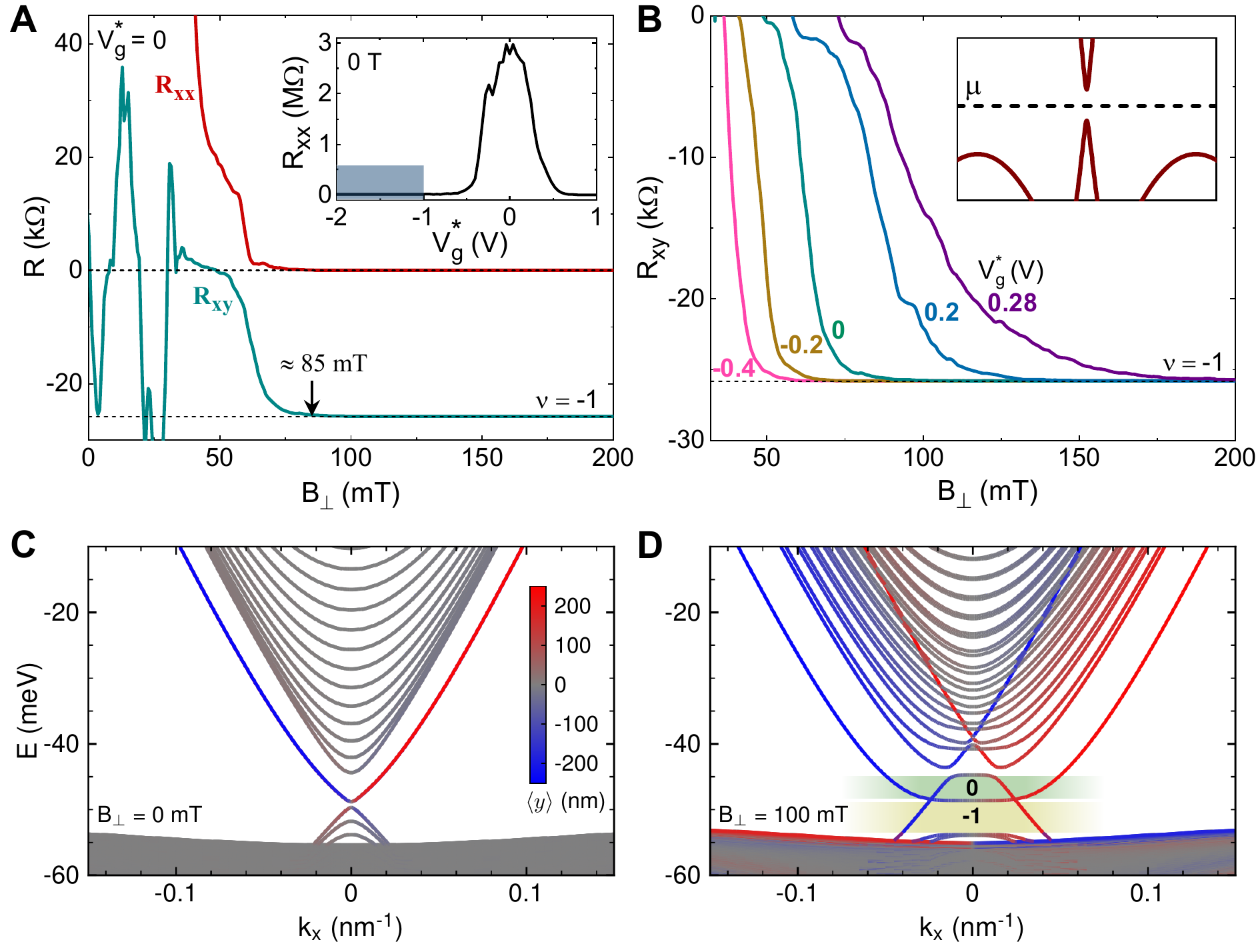}
\caption{\textbf{Early onset of $\nu=-1$ QH plateau in a (Hg,Mn)Te quantum well.}
(\textbf{A}) The longitudinal ($R_{xx}$) and transverse resistance ($R_{xy}$) of Dev 1 are depicted as a function of perpendicular magnetic field $B_\perp$ at $20$\,mK for normalized gate voltage $V_g^*=V_g-V_d=0$\,V. The inset shows the resistance $R_{xx}$ as a function of $V_g^*$ at $20$\,mK and $B_\perp=0$\,mT. The shaded region corresponds to the bulk $p$-regime.
(\textbf{B}) $R_{xy}$ as a function of $B_\perp$ for $V_g^*$ ranging from $-0.4$ to $0.28$\,V (near the QSH regime) at $20$\,mK for Dev 1. The dotted line shows the $\nu=-1$ QH plateau. The inset schematically shows the chemical potential $\mu$ in the bulk gap.
(\textbf{C}) Band structure computed using $\kdotp$ theory on a strip of width $500$\,nm (confinement in $z$- and $y$-direction) at $B_\perp=0$\,mT,
(\textbf{D}) Band structure for $B_\perp=100$\,mT. The expectation values $\avg{y}$ (extracted from eigenstates), represent the in-plane `location' of the state. Edge states are shown in red and blue while bulk states in grey. The topological gap (filling factor $\nu=0$ regime) and the magnetic gap with $\nu=-1$ are shaded in green and yellow, respectively.
}
\label{FigEarlyOnset}
\end{figure*}


\subsection{Chemical potential in the bulk gap - QSH regime}

The devices can be tuned from $n$- to $p$-regime by applying more negative gate voltage $V_g$, as shown in the inset of Fig.~2A. The bulk gap is around $V_g^*=0$\,V, where $V_g^*=V_g-V_d$ is the normalized gate voltage with $V_d$ being the gate voltage at maximal longitudinal resistance $R_{xx}$.
For the QH studies in this article, we focus on macroscopic devices, where $R_{xx}$ does not reach the value of $h/2 e^2$ characteristic of an ideal QSH edge configuration~\cite{SchmidtEA2012,VayrynenEA2014,BendiasEA2018}.
Having the chemical potential in the bulk gap, we observe that an out-of-plane magnetic field $B_\perp$ leads to an unexpected quantization of the transverse resistance $R_{xy}$ to $-h/e^2$ corresponding to a $\nu=-1$ QH plateau. Simultaneously, $R_{xx}$ drops to zero (Fig.~2A). The transition to the $\nu=-1$ QH plateau occurs at an anomalously low value of $B_\perp\approx 85$\,mT. This behaviour is unexpected from previous experimental investigations of undoped HgTe QWs in the QSH regime, where a transition to a $\nu=-1$ QH plateau was only observed for $B_\perp > 1.0 \, \mathrm{T}$~\cite{KonigEA2007}.
Theoretical investigations on the behaviour of QSH edge states in magnetic fields have shown that they can survive up to a few tesla, even in absence of protection by time reversal symmetry~\cite{TkachovHankiewicz2010,BenediktEA2012PRB,ChenEA2012PRB}. Chen \textit{et al.}~\cite{ChenEA2012PRB} have predicted `QAH-like' states in HgTe QWs with $\nu=\pm1$ in magnetic-field induced gaps. However, that analysis from the BHZ model suggests an onset field of the $\nu=-1$ plateau at $\sim 1$\,T, clearly larger than the onset field that we observe.

From an Arrhenius plot of conductance as a function of temperature, we estimate a bulk gap of $\sim 4.6$\,meV for Dev 1 (Supplementary Fig.~S1), which agrees well with the theoretical value of $4.0$\,meV. If the magnetization of Mn could close the topological bulk gap for one of the edge states \cite{LiuEA2008PRL101}, a $\nu=-1$ plateau would indicate a QAH phase. However, in our case, a magnetic field of $85$\,mT, corresponding to a spin polarization $\langle m \rangle \sim 0.15$ (Supplementary Eq.~\ref{EqnMagnetization}), can only close a gap of $\sim 1$\,meV~\cite{Boettcher19}. For this estimate, the combined effect of exchange interaction, Zeeman effect, and an additional orbital contribution has been taken into account \cite{Boettcher19}. Since the topological bulk gap is a factor of $\sim 5$ larger than this theoretical estimate, it is unlikely that the low-field $\nu=-1$ plateau in our (Hg,Mn)Te QW indicates a transition to the QAH phase.


\begin{figure}[!t]
\includegraphics[width=0.9\linewidth]{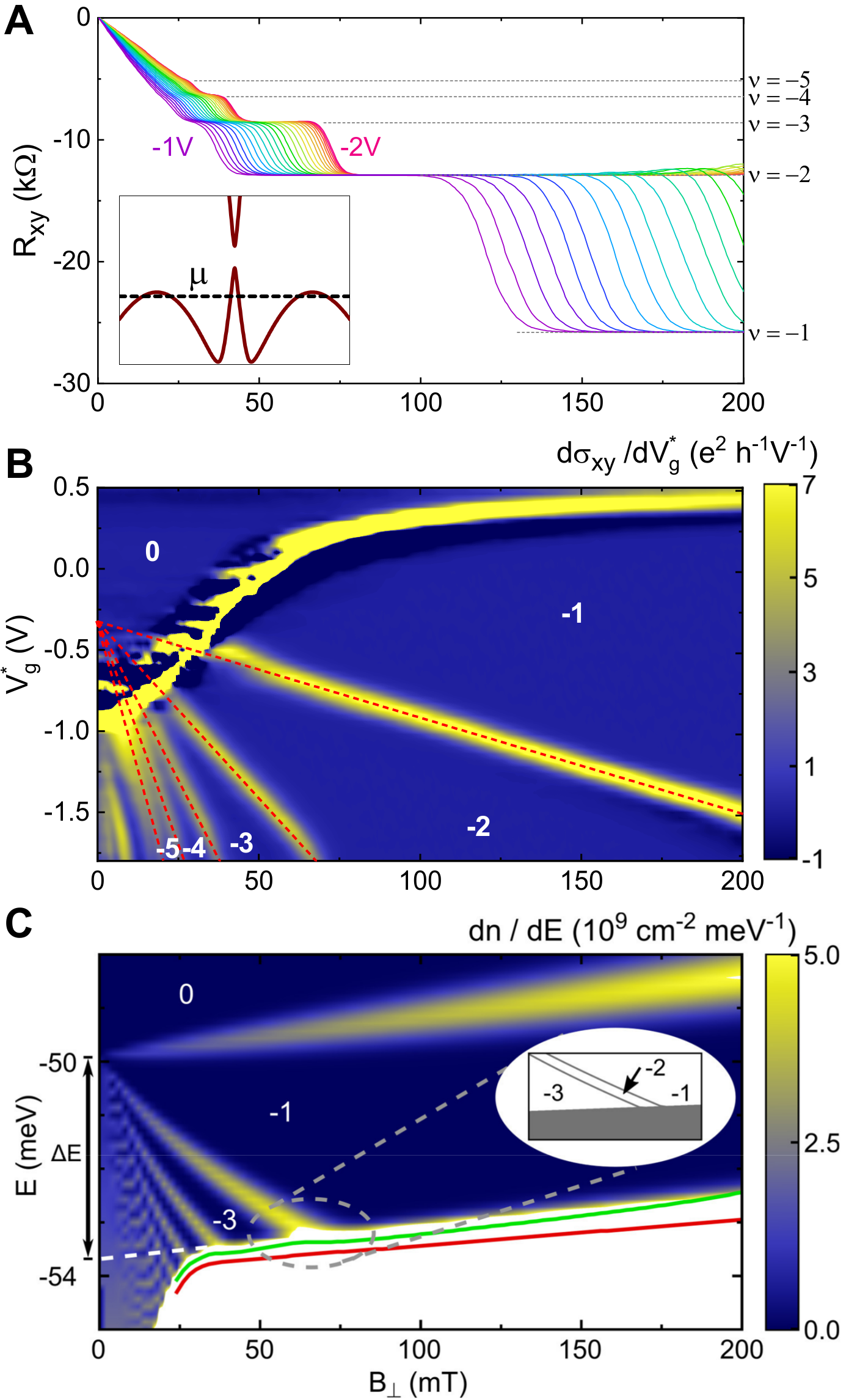}
\caption{\textbf{Higher Landau levels in bulk $p$-regime.}
(\textbf{A}) Transverse resistance $R_{xy}$ as a function of $B_\perp$ for $V_g^*$ ranging from $-1$ to $-2$\,V (Fig.~2A, inset, shaded area) at $20$\,mK for Dev 1. The dotted lines show the expected QH plateaus for $\nu=-1$ to $-5$. The inset shows schematically the chemical potential $\mu$ in the bulk $p$-regime.
(\textbf{B}) Experimental LL fan depicts $d\sigma_{xy}/dV_g^*$ as a function of $B_\perp$ and $V_g^*$ at $20$\,mK, where $\sigma_{xy}$ is the transverse conductivity.  Numbers indicate QH filling factors $\nu$, while the red dashed lines show transitions between adjacent LLs. The label `$0$' indicates the quantum spin Hall state (edge states not shown).
(\textbf{C}) Theoretical LL fan chart for a QW including the first 1000 LLs. Colour code indicates DOS, $dn/dE$, where white implies that the DOS is out-of-scale. The green and red curves correspond to a chemical potential for densities $p=5\times10^9$\,$\mathrm{cm^{-2}}$ and $2\times 10^{10}$\,$\mathrm{cm^{-2}}$, respectively. The white dashed line extrapolates the camelback edge to $B_\perp = 0$; $\Delta E$ is the energetic difference between the valence band edge and the camelback. The inset shows a zoom of the lowest LLs without the effect of broadening.
}
\label{FigHigherLL}
\end{figure}


The early onset of the $\nu=-1$ QH plateau occurs for $V_g^*$ ranging from $0.28$ to $-0.4$\,V with the onset field as low as $50$\,mT (Fig.~2B). We emphasize that the $\nu=-1$ plateau (near the QSH phase) cannot occur from the conventional QH effect in a low density system for the following reasons. Firstly, we observe the $\nu=-1$ plateau even for small positive $V_g^*$ (Fig.~2B) where we would conventionally expect a $\nu=+1$ plateau. Secondly, in this picture we would expect the $p$-density to increase for more negative $V_g^*$, which would imply higher fields to observe the QH plateaus. This is in direct contrast to Fig.~2B, where we observe lower onset fields for more negative $V_g^*$.  The occurrence of QH plateaus at such low magnetic fields has not been reported so far.

To find out the microscopic origin of the low-field $\nu=-1$ QH plateau, we have calculated dispersions in the presence of a magnetic field. Figure~2C,D show $\mathbf{k}\cdot\mathbf{p}$ band structures of a (Hg,Mn)Te sample in a strip geometry. The width $w=500$\,nm is smaller compared to that of the measured devices for practical reasons, but this choice will not affect our conclusions.  Energies $E$ are defined with respect to the conduction band ($\Gamma_8$) of unstrained bulk HgTe (Supplementary Section~S4). For zero magnetic field (Fig.~2C), we can clearly distinguish the bulk bands (grey) and the edge states (red and blue).  The (two-fold degenerate) edge states cross (up to a small finite size gap) at the top of the valence band at the $\Gamma$ point. Below $E\approx -52$\,meV, there is a high density of bulk valence band states, brought about by the camelback structure. The top of the valence band is at $k_x\approx 0.4$\,nm$^{-1}$, outside the plot range.

Upon application of a magnetic field, the different values of the magnetic gauge field $\mathbf{A}$ at the opposite edges (recall the Peierls substitution $k_x\to k_x + eA_x/\hbar$) cause the degeneracy of the edge states to be lifted. The crossing at $k_x=0$ splits into two copies, one moving up, and one moving down in energy (Fig.~2D). The lower energy copy disappears rapidly due to hybridization of the edge states with the bulk. The higher energy copy survives, even in the bulk conduction band.
Hybridization between these edge states and the bulk states is (almost) absent, because they differ in orbital character as well as in wave function localization.
Simultaneously, LLs form, starting near the original (zero-field) Dirac point near $E=-49$\,meV. The topological gap, characterized by a pair of counterpropagating edge states, is reduced in magnetic fields (see Fig.~2D, green shading)~\cite{TkachovHankiewicz2010,Boettcher19}. Below this energy, the single edge state survives indicating a transition to the magnetic gap (gap between LLs induced by the magnetic field; see Fig.~2D, yellow shading). In contrast to the QAH effect, where a $\nu=-1$ plateau arises in the topological gap as a result of a bulk band gap closing (for one `spin' channel) driven by the magnetization of Mn \cite{LiuEA2008PRL101}, the present $\nu=-1$ plateau appears in a magnetic gap in the absence of a bulk band gap closing. The pre-existence of the QSH edge state leads to the emergence of a QH state at effectively lower magnetic fields.

\subsection{Chemical potential in the bulk $p$-regime}

The early onsets of the emergent QH plateaus also occur when the chemical potential is tuned into the bulk $p$-regime (schematic in the inset of Fig.~3A). In this regime, we see QH plateaus from $\nu=-1$ to $ -5$ at all values of $V_g^*$ from $-1$ to $-2$\,V.
The well-resolved QH plateaus are visible in $R_{xy}$ measured as a function of $B_\perp$ (Fig.~3A) and in the LL fan (Fig.~3B). All the QH plateaus are observed at remarkably low values of $B_\perp < 150$\,mT, while at certain gate voltages, resolved plateaus can be seen
for $B_\perp$ as low as $20$--$30$\,mT. Additionally, the $\nu=-1$ plateau persists up to exceptionally high $B_\perp \sim 9$\,T (Supplementary Fig.~S3).
The origin of these QH plateaus is different from those in the previous section, as we will explain below.


\begin{figure*}[tb]
\includegraphics[width=1\linewidth]{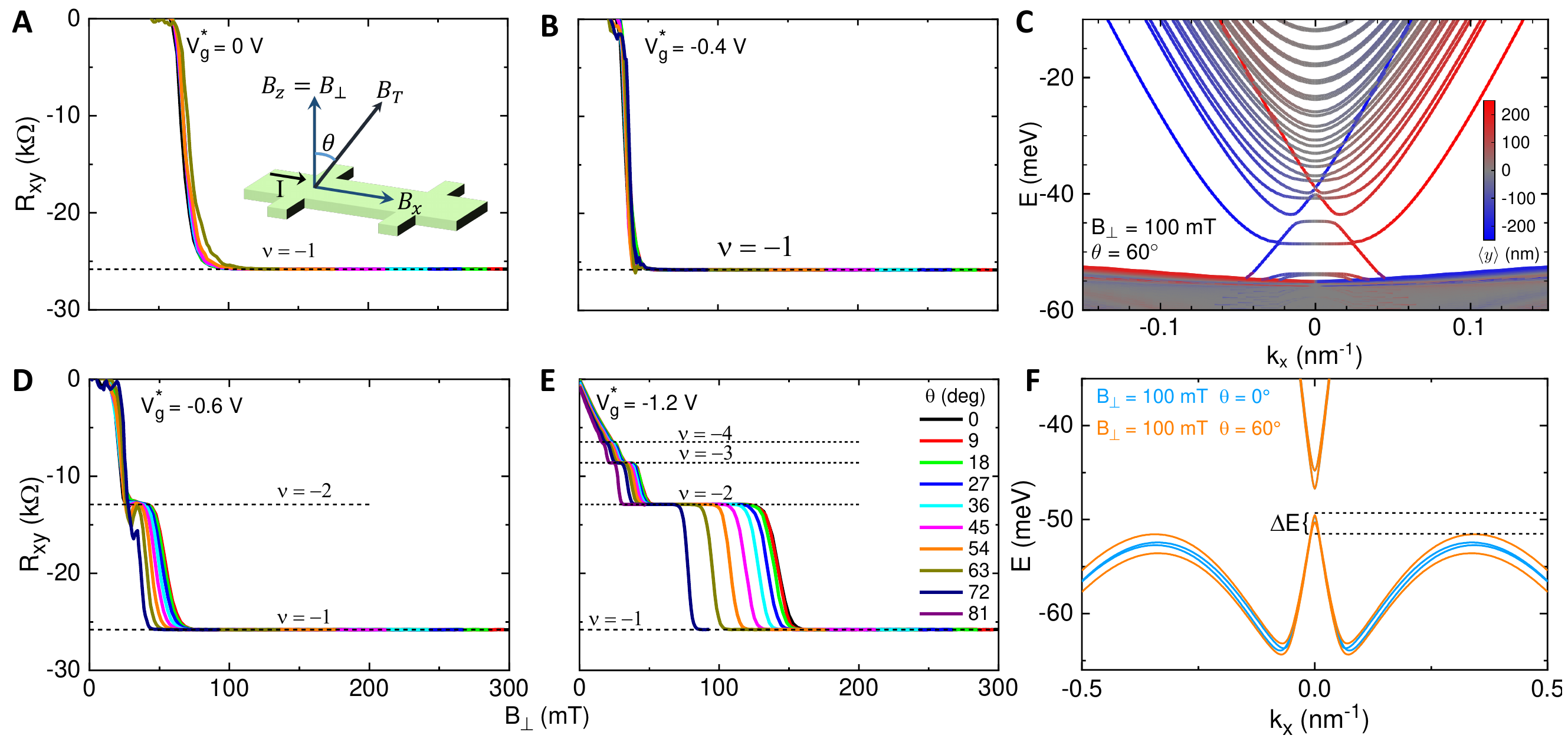}
\caption{\textbf{Effect of in-plane magnetic field on quantum Hall plateaus.} The transverse resistance $R_{xy}$ as a function of the out-of-plane component of the magnetic field $B_\perp=B_z$  for various values of the angle $\theta$ at
(\textbf{A}) $V_g^*=0$\,V, and
(\textbf{B}) $V_g^*=-0.4$\,V.
The magnetic field components are $\vec{B}=(B_x,B_y,B_z)=(B_\parallel,0,B_\perp)=B_T(\sin\theta,0,\cos\theta)$.
The inset in \textbf{A} shows the measurement scheme with the Hall bar in the $x$-$y$ plane. The current $I$ flows along the length of the Hall bar, while the total magnetic field $B_T$ is applied at angle $\theta$ to the $z$-direction.
(\textbf{C}) Band structure computed on a strip of width $500$\,nm for $\vec{B}$ at an angle $\theta=60^{\circ}$ such that $B_\perp=100$\,mT.
$R_{xy}$ as a function of $B_\perp$  for various values of $\theta$ at
(\textbf{D}) $V_g^*=-0.6$\,V, and
(\textbf{E}) $V_g^*=-1.2$\,V.
The chemical potential is in the gap regime for \textbf{A}, \textbf{B} and in the $p$-regime for \textbf{D}, \textbf{E}.
(\textbf{F}) At constant $B_\perp = 100 \,\mathrm{mT}$, the effect of an in-plane magnetic field on the band structure is depicted. An in-plane magnetic field affects mainly the height of the camelback, but not the valence band states near $k=0$. $\Delta E$ (see also Fig.~3C) shows the difference between the energy at the camelback and $k=0$.}
\label{FigRotation}
\end{figure*}


To understand these low-field QH plateaus, we show in Fig.~3C the calculated bulk LL fan for an infinite QW in the $p$-regime; the colour code indicates the DOS.
In the following, we assume that the DOS of each LL exhibits a Gaussian broadening as used in Ref.~\cite{NovikEA2005}. In Fig.~3C, the white area in the valence band ($E<-52 \, \mathrm{meV}$) indicates that the DOS is $\sim 400$ times larger than in a single, conventional LL. This regime is formed by a very dense collection of LLs which arise due to the camelback. In particular, below $B_\perp < 70 \, \mathrm{mT}$, LLs with small LL indices ($j=-2,-1,\ldots $), resulting from highly mobile carriers at small $k$, can coexist at the same energy with higher LLs ($j \sim 500$--$1000$), stemming from the camelback.   For $V_g^*<-0.6 \, \mathrm{V}$, the chemical potential is pinned to the upper edge of the camelback since any small decrease of the chemical potential would result in a large increase of the $p$-density. Therefore, the large density of states  originating from the camelback (van Hove singularity) is responsible for an early onset of QH plateaus with $\nu =-1$ to $-5$ for a large range of $p$-densities,  as demonstrated in Fig.~3C  (see also Supplementary Section~S6). This finding is in good qualitative agreement  with the experimental results (Fig.~3A,B). The small size of even plateaus in the theoretical result is a side effect of neglecting the effects of bulk-inversion asymmetry terms for which the exact strength is not known for (Hg,Mn)Te QWs (Supplementary Section~S6 and Fig. S10). In addition, pinning to the camelback can also explain the exceptionally long $\nu=-1$ QH plateau which is observed in the experiment (Supplementary Section~S1).

Another consequence of the pinning mechanism is that all observed QH plateaus in the $p$-regime are very sensitive to the difference in energy of the valence band at $k=0$ and the maximum of the camelback at $B_\perp=0$ ($\Delta E$ in Fig.~3C,4F): The onset fields of the QH plateaus decreases as $\Delta E$ decreases. This results from the fact that the upper edge of the camelback in magnetic fields (dashed white line in Fig.~3C) extrapolates to the camelback at $B_\perp=0$ (cf.\ Fig.~1B and Fig.~3C).  Consequently, we expect that any (Hg,Mn)Te TI with a direct gap and close to the direct-indirect-gap transition (dashed line in Fig.~1A) exhibits a similar characteristic behaviour in the $p$-regime. In contrast to the Mn-free case, we expect that for Mn-doped QWs the onset fields are shifted to even smaller $B_\perp$, since the exchange coupling increases the slope of the lowest LLs, and the camelback height in magnetic fields.

Despite the macroscopic occupation of bulk states near the camelback, the bulk conduction remains suppressed at low temperatures, because bulk carriers are localized by disorder. Indeed, the higher Landau plateaus are well resolved in experiments at $20$\,mK, but they are not robust against increased temperatures (Supplementary Section~S1B,D). The mobile $p$-carrier density as estimated from the classical Hall slope at low field ($B_\perp < 20 \; \mathrm{mT}$) is $\sim 2$--$3 \times 10^9\,\mathrm{cm}^{-2}$ for the entire range of $V_g^*$ shown in Fig.~3A (see also Supplementary Section~S2) which agrees well with the theoretically calculated density of the highly mobile carriers lying above the camelback ($p\approx2\times10^9 \; \mathrm{cm}^{-2}$). The density of localized carriers at the camelback as estimated from the known gate capacitance is $4$--$5 \times 10^{11}\,\mathrm{cm}^{-2}$. These are significant for efficiently screening the disorder and are thus leading to highly mobile carriers. As extracted from the Drude model and the experimentally measured mobile carrier density and conductivity, the $p$-carrier mobility is $\sim 6$--$9 \times 10^5\,\mathrm{cm}^{2}\,\mathrm{V}^{-1}\,\mathrm{s}^{-1}$ (Supplementary Section~S2). Thus, the van Hove singularity at the camelback is crucial for observing perfectly quantized Hall plateaus at such low carrier densities and anomalously low magnetic fields.

\subsection{Magneto-transport in tilted magnetic fields}

To further distinguish the magneto-transport in the QSH and in the bulk $p$-regime, we have performed measurements in tilted magnetic fields. The measurement schematic is shown in the inset of Fig.~4A where the sample is in the $x$-$y$ plane and the magnetic field $\mathbf{B}=(B_x,B_y,B_z)=(B_\parallel,0,B_\perp)=B_T(\sin\theta,0,\cos\theta)$ is applied at an angle $\theta$ to the $z$-direction. The measurements are performed for various values of $\theta$. The current $I$ flows along the length of the sample. We choose four values of $V_g^*$ which encompass the two regimes: two points in the vicinity of the QSH regime (chemical potential in the bulk gap) and two in the bulk $p$-regime. In the QSH regime, the onset field for the $\nu=-1$ QH plateau depends only on $B_\perp=B_z$ (shown for $V_g^*=0$ and $-0.4$\,V in Fig.~4A,B, respectively) and not on $B_x$ for all values of $\theta$. This is markedly different from the $p$-regime, where the onset fields for all observed QH plateaus  with $\nu=-1$ to $-5$ are strongly dependent on $\theta$. For larger $\theta$, all QH plateaus (including $\nu=-1$) appear at lower $B_\perp$ as shown for $V_g^*=-0.6$ and $-1.2$\,V in Fig.~4D,E, respectively.

The observed differences can be captured within our theoretical model. Figure~4C shows the dispersion for a strip geometry with the magnetic field at an angle $\theta=60^{\circ}$ and $B_\perp=100$\,mT.
Comparing this band structure to Fig.~2D, we find that an in-plane magnetic field does not alter the band structure close to $k_x=0$ significantly.
This agrees well with the experimental observation in Fig.~4A,B indicating that the $\nu=-1$ plateau is indeed related to the QH state forming outside of the topological gap. In contrast, the height of the camelback is very sensitive to the in-plane magnetic field and increases for larger $\theta$ (implying that $\Delta E$ decreases), as shown in Fig.~4F. The decrease of $\Delta E$   indicates lower onset fields for the QH plateaus in the  bulk $p$-regime, as observed experimentally in Fig.~4D,E.
The difference between the $k=0$ and the camelback regime is connected to the interplay between the exchange coupling and the variation in orbital character of the band structure as a function of momentum (Supplementary Section~S5). A strong in-plane field dependence of the camelback can therefore only exist in Mn-doped samples.

We do not see any evidence of early onsets of QH plateaus for Dev 2. The QH plateaus for Dev 2 occur for $B_\perp > 1$\,T (Supplementary Fig.~S6b). Since Dev 2 has an indirect gap, this observation is in agreement with our model which predicts an early onset of QH plateaus only for direct gap TIs close to the direct-indirect transition (Fig.~1A). This also highlights the crucial role of Mn in tuning the band structure and in realizing the emergent quantum Hall phenomena, since the two devices have the same thickness and differ only in Mn concentration.

\section{Summary and Outlook}

To conclude, we have found two different regimes with emergent QH states in (Hg,Mn)Te quantum wells. When the chemical potential is in the band gap, a $\nu=-1$ QH state forms from one of the QSH edge states at unusually small perpendicular magnetic fields. This QH state is observed only for samples with an inverted band structure indicating the crucial role of topological edge channels. In contrast, the QH plateaus observed when the chemical potential is in the bulk $p$-regime have a very different origin compared to the  $\nu=-1$ plateau observed in the QSH regime. They form due to the large DOS related to the camelback (the van Hove singularity). The emergence of QH plateaus at such low fields provides a prospective platform to realize chiral Majorana fermions when integrated with superconductors. Since the emergent quantum Hall states are inherently connected to the inverted band structure of HgTe, it should lead to further experimental and theoretical investigation of these effects in other two dimensional topological materials such as InAs/GaSb double quantum wells and $\mathrm{WTe_2}$.

\section{Materials and Methods}
\textbf{Material growth and characterization}:
The (Hg,Mn)Te QWs investigated in this work are grown by molecular beam epitaxy on commercial (Cd,Zn)Te substrates. A $50$\,nm CdTe buffer layer is grown on the substrate. The (Hg,Mn)Te QWs are sandwiched between (Hg,Cd)Te bottom and top barriers of thickness $\sim 150$\,nm and $\sim 50$\,nm, respectively. The thickness of the QW and Mn concentration are confirmed by X-ray diffraction measurements.

\textbf{Device fabrication}:
The QWs are fabricated into Hall bars using optical lithography and conventional dry etching (using accelerated argon ions).  Ohmic contacts are formed by deposition of $50$\,nm AuGe and $50$\,nm Au by e-gun evaporation. A $110$\,nm of dielectric layer (alternating layers of $10$\,nm $\mathrm{SiO_2}$ and $10$\,nm of $\mathrm{Si_3N_4}$) was grown by plasma enhanced chemical vapour deposition. The carrier density can be tuned by applying voltage to a $200$\,nm metallic top gate (Ti/Au) deposited on the dielectric insulator.

\textbf{Magneto-transport measurements}:
All magneto-transport measurements have been performed in a dilution refrigerator of base temperature $20$\,mK and equipped with a 3D vector magnet. We have used four-probe ac low frequency ($\sim 13$\,Hz) measurement techniques employing lock-in amplifiers to measure the longitudinal and transverse resistance, $R_{xx}$ and $R_{xy}$ respectively, of the devices. The current flowing through the device was measured continuously by measuring the voltage across a standard resistor and was in the range $2$--$20$\,nA.

\subsection*{Acknowledgements}

We acknowledge discussions with S.-C. Zhang, B. A. Bernevig, C.-X. Liu, X.-L. Qi, C. Gould, and C. Br\"{u}ne. We thank K. Bendias for fabricating some of the devices. We acknowledge financial support from the Deutsche Forschungsgemeinschaft (DFG, German Research Foundation) in the Leibniz Program and in the projects SFB 1170 (Project ID 258499086) and SPP 1666 (Project ID 220179758), from the EU ERC-AdG program (Project 4-TOPS), from the W\"{u}rzburg-Dresden Center of Excellence `Complexity \& Topology in Quantum Matter' (CT.QMAT), and from the Free State of Bavaria (Elitenetzwerk Bayern IDK `Topologische Isolatoren' and the Institute for Topological Insulators).

\subsection*{Author contributions}

L.W.M. and H.B. planned the project. S.S. conducted the measurements and analyzed the data with help from A.B. and P.S.  The band structure calculations were performed by W.B. and J.B. The material was grown by P.L. and L.L. The device was fabricated by P.S. All authors contributed to interpretation of the results. L.W.M., H.B. and E.M.H. supervised the project. All authors participated in writing the manuscript led by S.S., W.B., and J.B.

\subsection*{Competing interests}

The authors declare no competing interests.

\subsection*{Data and materials availability}

All data needed to evaluate the conclusions in the paper are present in the paper and/or the Supplementary Materials. Additional data available from authors upon request.

\onecolumngrid
\newpage
\begin{center}
{\noindent\large\bf Supplementary Materials for\\  \emph{Emergent quantum Hall effects below $50$\,mT in a two-dimensional topological insulator}}
\vspace{1em}
\end{center}

\setcounter{page}{1} \renewcommand{\thepage}{S\arabic{page}}

\setcounter{figure}{0}   \renewcommand{\thefigure}{S\arabic{figure}}

\setcounter{equation}{0} \renewcommand{\theequation}{S.\arabic{equation}}

\setcounter{table}{0} \renewcommand{\thetable}{S.\arabic{table}}

\setcounter{section}{0} \renewcommand{\thesection}{S\arabic{section}}

\renewcommand{\thesubsection}{S\arabic{section}.\Alph{subsection}}


\makeatletter
\renewcommand*{\p@subsection}{}
\makeatother

\renewcommand{\thesubsubsection}{S\arabic{section}.\Alph{subsection}-\arabic{subsubsection}}

\makeatletter
\renewcommand*{\p@subsubsection}{}  
\makeatother



\noindent The supplementary information includes the following:\\

\noindent Section S1. Further experimental data for Dev 1\\
Section S2. Estimation of carrier density and mobility: mobile and localized carriers\\
Section S3. Magneto-resistance for Dev 2\\
Section S4. $\kdotp$ calculations for a strip geometry\\
Section S5. Model for magnetic response of manganese\\
Section S6. Further theoretical data for Dev 1 and Dev 2\\
Fig. S1. Determination of band gap of Dev 1.\\
Fig. S2. Magnetoresistance for Dev 1.\\
Fig. S3. Quantum Hall measurements for Dev 1.\\
Fig. S4. Temperature dependence of the onset field for quantum Hall plateaus for Dev 1.\\
Fig. S5. Density and mobility of charge carriers.\\
Fig. S6. Magnetoresistance for Dev 2.\\
Fig. S7. Band structure calculations for Dev 1.\\
Fig. S8. Band structure calculations on a strip for Dev 2.\\
Fig. S9. Band structure calculations for Dev 2.\\
Fig. S10. Influence of bulk inversion asymmetry for Dev~1.\\
References \cite{Kane1957,BerdingEA2000,ARPACK,GuiEA2004,Winkler2003_book}

\newpage

\section{Further experimental data for Dev 1}

\subsection{Experimental determination of band gap}
The bulk band gap has been determined from an Arrhenius plot of maximum resistance (at the charge neutrality point) as a function of temperature. We measure the longitudinal resistance $R_{xx}$ as a function of normalized gate voltage $V_g^* = V_g-V_d$, where $V_d$ corresponds to the gate voltage for which $R_{xx}$ is maximum, for different temperatures $T$ (Fig.~S1a for $T=15$\,K). The minimum conductance $G_\mathrm{min}$ is evaluated as $G_\mathrm{min}=1/{R_{xx}^\mathrm{max}}$, where $R_{xx}^\mathrm{max}$ is the maximum resistance in the $R_{xx}$-$V_g$ curve. The band gap can then be evaluated by fitting the function
\begin{equation}\label{EqnArrhenius}
  G_\mathrm{min}=G_0\exp{\left(-\frac{E_G}{2\kB T}\right)},
\end{equation}
to the $G_\mathrm{min}$-$T$ curve (Fig.~S1b). This fitting yields a band gap $\sim 4.5$\,meV for Dev 1.


\begin{figure}[bth]
\includegraphics[width=0.8\linewidth]{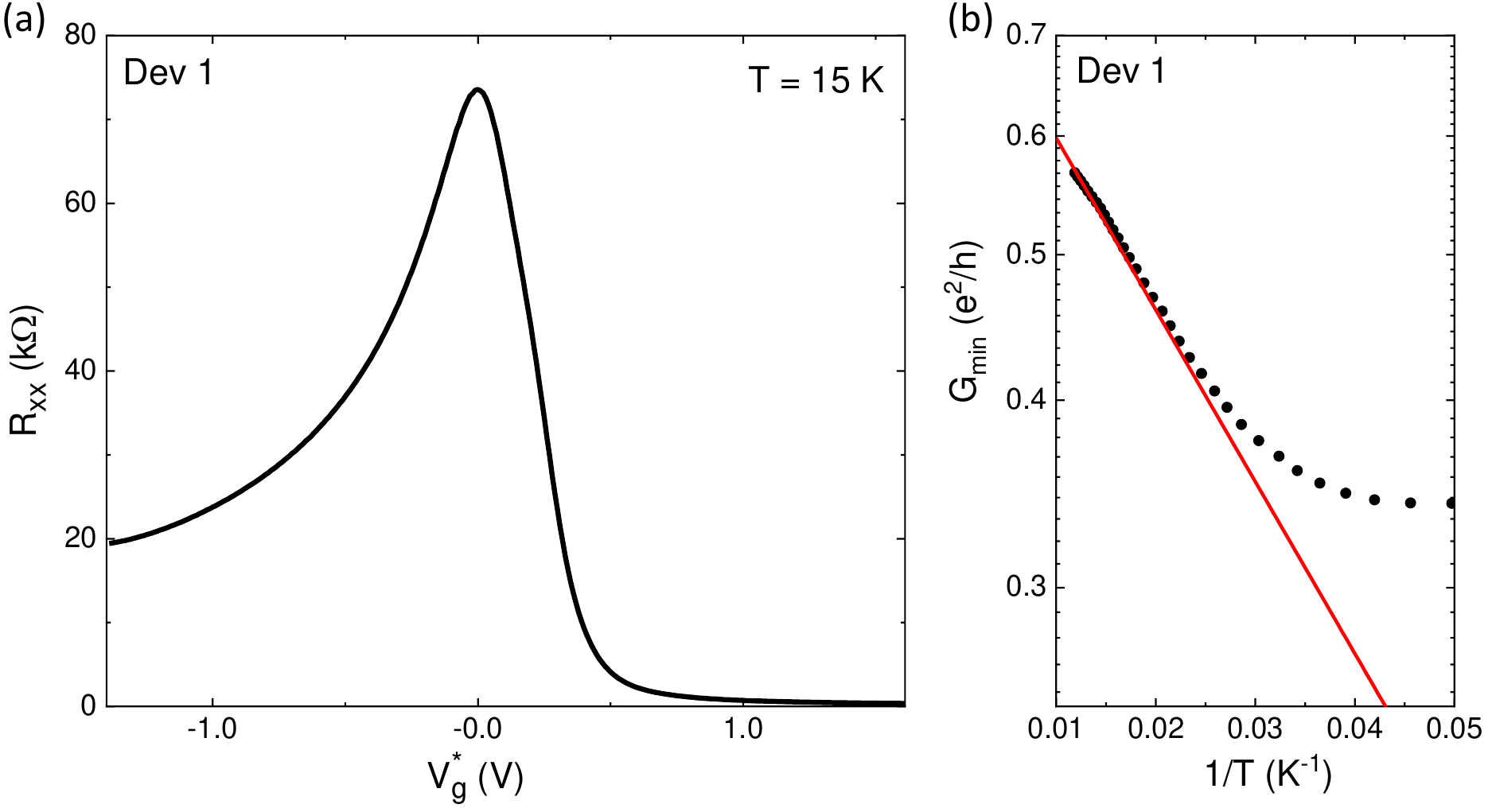}
\caption{\textbf{Determination of band gap of Dev 1.}
(a) The longitudinal resistance $R_{xx}$ of Dev 1 as a function of normalized gate voltage $V_g^*$ at temperature $T=15$\,K.
(b) The minimum conductance $G_\mathrm{min}$ as a function of $1/T$. The red line shows fit according to Eq.~(\ref{EqnArrhenius}).}
\label{FigBandGap}
\end{figure}


\subsection{Longitudinal resistance $R_{xx}$ for Dev 1}

Figure~S2 shows the longitudinal and transverse resistance, $R_{xx}$ and $R_{xy}$ respectively, as a function of perpendicular magnetic field $B_\perp$ for $V_g^*=-1$\,V (see Fig.~S1 and inset of Fig. 2A in the main text) at $20$\,mK. $R_{xx}$ shows Shubnikov-de~Haas oscillations and $R_{xy}$ shows quantum Hall plateaus. For filling factors $\nu=-1$ and $-2$, we find that $R_{xx}=0$.


\begin{figure}[!h]
\includegraphics[width=0.82\linewidth]{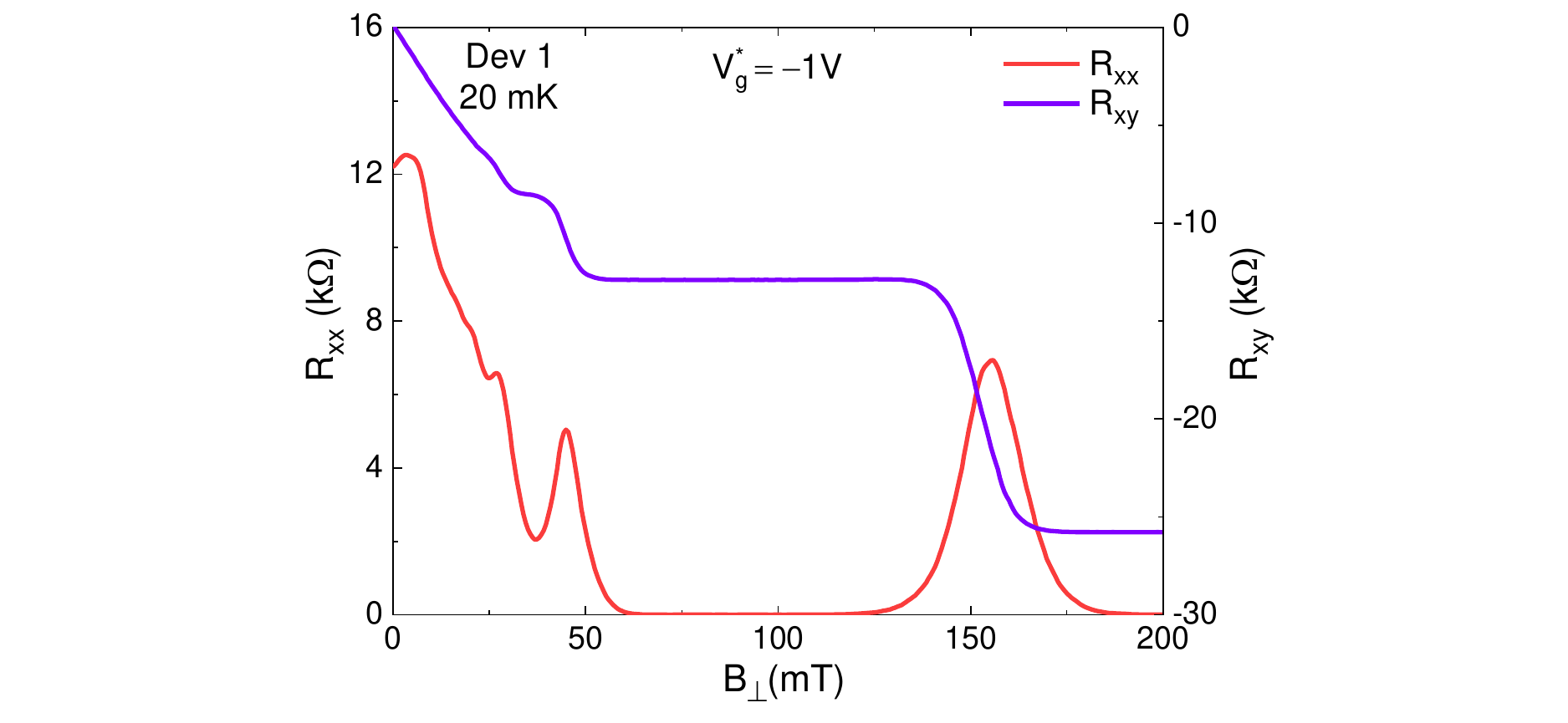}
\caption{\textbf{Magnetoresistance for Dev 1.} The longitudinal and transverse resistance, $R_{xx}$ and $R_{xy}$ respectively, as a function of perpendicular magnetic field $B_\perp$ for normalized gate voltage $V_g^*=-1$\,V (see Fig.~S1 and inset of Fig. 2A in the main text) at temperature $T=20$\,mK.
}
\label{FigMR_Rxx_Dev1}
\end{figure}


\subsection{Quantum Hall measurements for Dev 1}


\begin{figure}[!h]
\includegraphics[width=1\linewidth]{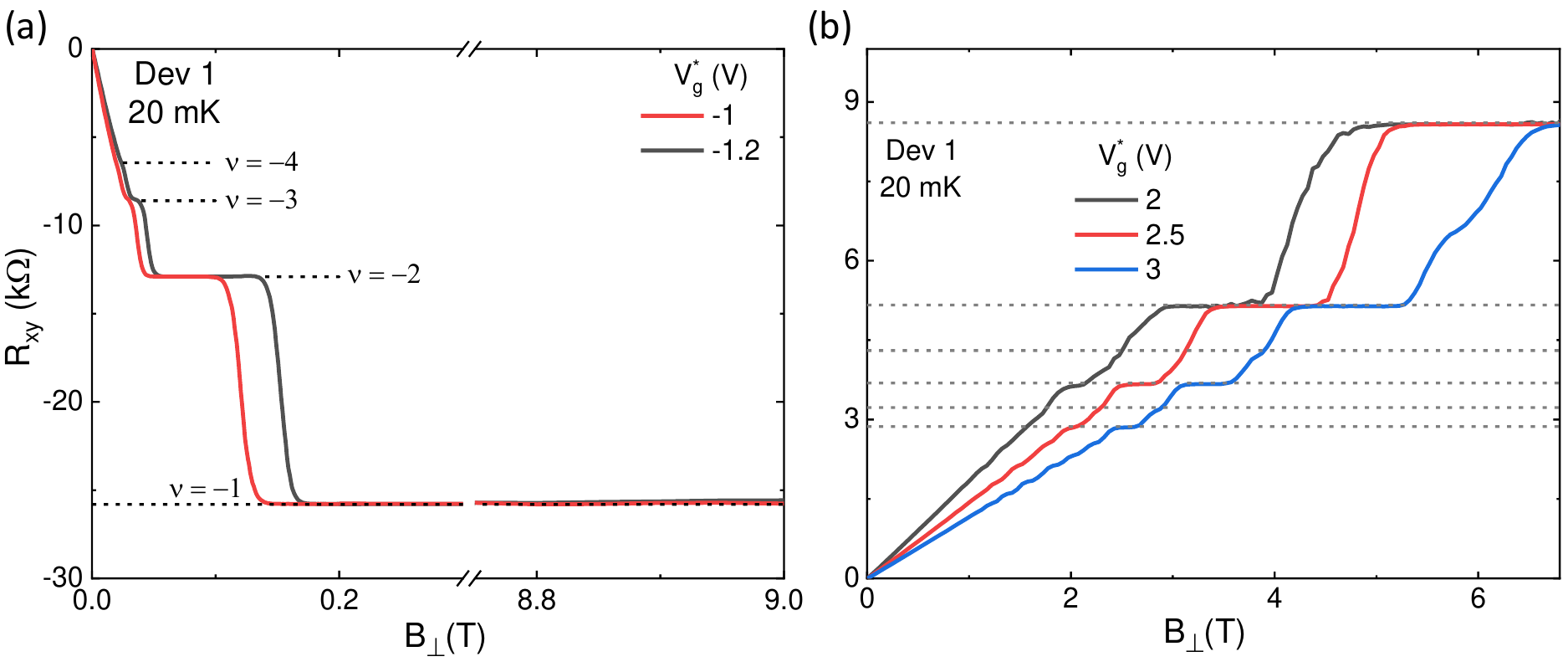}
\caption{\textbf{Quantum Hall measurements for Dev 1.} (a) The transverse resistance $R_{xy}$ as a function of perpendicular magnetic field $B_\perp$ for normalized gate voltage $V_g^*=-1$ and $-1.2$\,V at temperature $T=20$\,mK. The $\nu=-1$ plateau is exceptionally long and persists up to $9$\,T. (b) $R_{xy}$ as a function of $B_\perp$ for $V_g^*=2$, $2.5$ and $3$\,V at $T=20$\,mK. The dashed gray lines indicate quantized Hall resistance in units of $h/\nu e^2$, where $\nu$ is the filling factor.
}
\label{FigLongPlateau}
\end{figure}


\subsubsection{Exceptionally long $\nu=-1$ QH plateau in the $p$-doped regime}

Figure~S3a shows $R_{xy}$ as a function of $B_\perp$ for $V_g^*=-1$ and $-1.2$\,V at $20$\,mK. The $\nu=-1$ plateau is exceptionally long and persists up to $9$\,T which cannot be explained by conventional QH physics of a 2D system. The exceptionally long plateau can be explained by the pinning mechanism (see main text and Supplementary Section~\ref{SuppSecTheoryDev1Dev2}).

\subsubsection{Quantum Hall plateaus in the $n$-doped regime}

Figure~S3b shows $R_{xy}$ as a function of $B_\perp$ for $V_g^*=2$, $2.5$ and $3$\,V at $20$\,mK. The device is in the $n$-doped regime and we observe the expected QH plateaus. There is no early onset of QH plateaus when the chemical potential is in the $n$-doped regime.

\clearpage

\subsection{Dependence of the onset fields on temperature for the emergent quantum Hall plateaus}
The early onset to a $\nu=-1$ QH plateau in the QSH regime (in the band gap) persists for temperatures up to $2$\,K with the onset field increasing for higher temperatures (Figs.~S4a and S4c).
The effect of temperature on the band structure is via the exchange interaction [Eq.~\eqref{EqnMagnetization}] and hence the magnetization of the Mn atoms. This leads to an increase in the magnetic field required to split the edge states to form the quantum Hall state.
In the bulk $p$-regime, the higher QH Hall plateaus ($\nu=-2,-3,-4$ and $-5$) are resolved only for the lowest temperature ($20$\,mK). As shown in Fig.~S4b, $\nu=-1, -2$ and a weak $-3$ QH plateau are observed for $T=280$\,mK. At $2$\,K, no QH plateaus are visible for $V_g^*=-1$\,V within the range of magnetic field investigated here.

\vspace{1 cm}


\begin{figure*}[th]
\includegraphics[width=0.85\linewidth]{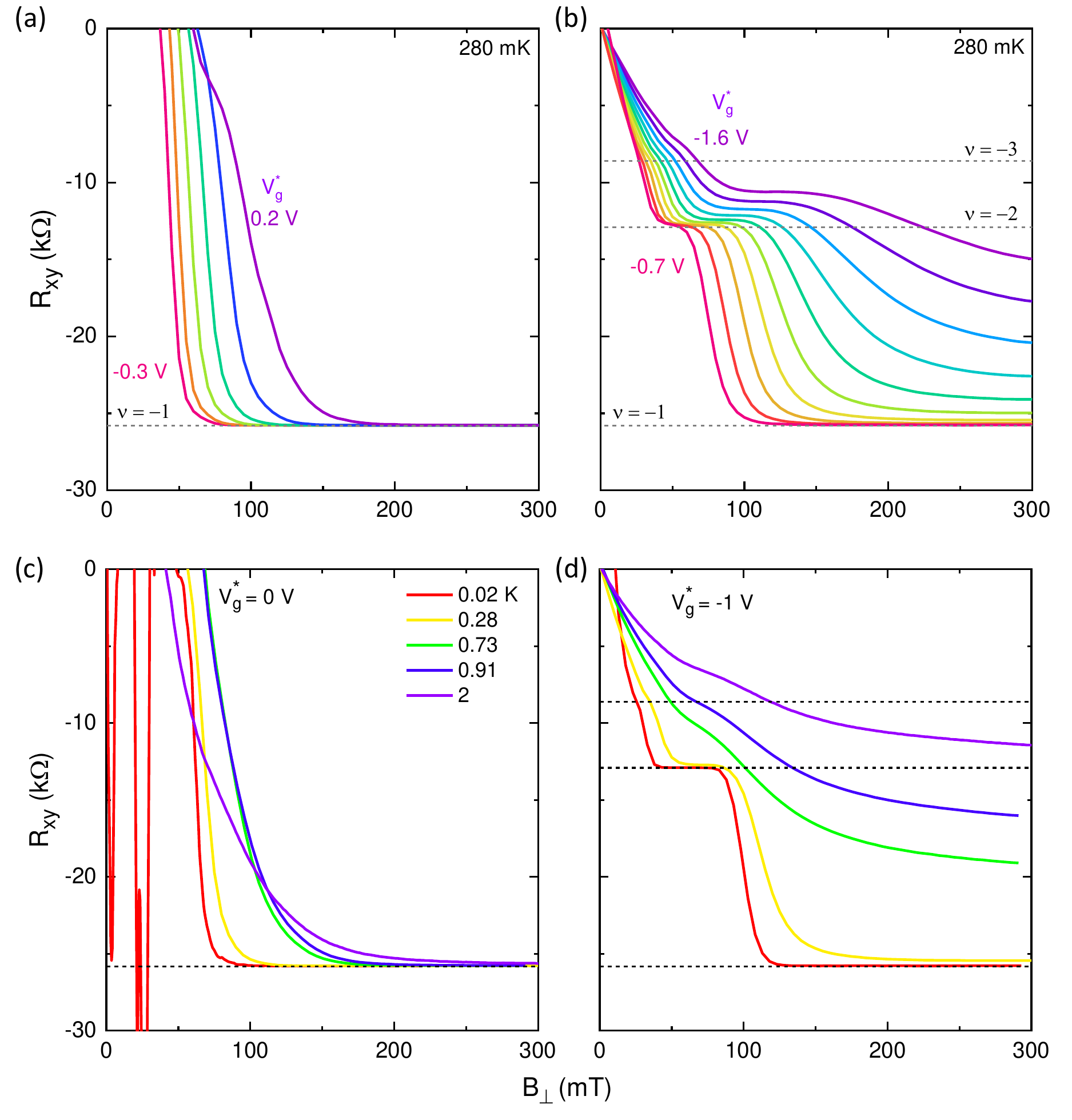}
\caption{\textbf{Temperature dependence of the onset field for quantum Hall plateaus for Dev 1.} The transverse resistance $R_{xy}$ as a function of the out-of-plane component of the magnetic field $B_\perp$ at
(a) $280$\,mK for $V_g^*$ ranging from $0.2$ to $-0.3$\,V.
(b) $V_g^*$ ranging from $-0.7$ to $-1.6$\, V.
(c) $V_g^*=0$\,V.
(d) $V_g^*=-1$\,V for different temperatures ranging from $0.02$ to $2$\,K.}
\label{FigTDep_Supp}
\end{figure*}


\clearpage

\section{Estimation of carrier density and mobility: Mobile and localized carriers}

The carrier density $n$ has been estimated as $n=1/R_{H}e$, where $R_H$ is the Hall coefficient which is calculated from the slope of the $R_{xy}$--$B_\perp$ curve (inset of Fig.~S5a,c) and $e$ is the electron charge. The carrier mobility $\mu$ is calculated from the Drude model as $\mu=\sigma_{xx}/n e$, where $\sigma_{xx}$ is the experimentally measured longitudinal conductivity.

Figure~S5a shows that $n$ is $1.5$--$3.5\times 10^{11}\,\mathrm{cm}^{-2}$ for $V_g^*$ ranging from $1$ to $2$\,V which corresponds to a gate efficiency of $2 \times 10^{11}\,\mathrm{cm}^{-2}\mathrm{V}^{-1}$ in the bulk $n$-regime. The maximum electron mobility is $\sim 1.5\times10^5\,\mathrm{cm}^2\mathrm{V}^{-1}\mathrm{s}^{-1}$ (Fig.~S5b). The estimation of hole density and mobility is more challenging because of the formation of quantum Hall plateaus at very low fields ($\sim 20$--$30$\,mT). However for $B_\perp<20$\,mT, $R_{xy}$ varies linearly with $B_\perp$ and hence the slope of the curve in this narrow window allows us to estimate the hole density. The hole density varies from $2$--$3 \times 10^{9}\,\mathrm{cm}^{-2}$ for the entire range of $V_g^*$ measured in the experiment (Fig.~S5c), which agrees well with the theoretically estimated value. This weak dependence of $n$ on gate voltage in the $p$-regime agrees with our pinning model as explained in the main text. The maximum hole mobility as calculated from the Drude model is $\sim 9 \times10^5\,\mathrm{cm}^2\mathrm{V}^{-1}\mathrm{s}^{-1}$ (Fig.~S5d). We emphasize that the numbers for hole density and mobility mentioned above correspond to mobile carriers. At lowest temperatures of $20$\,mK, the carriers near the camelback are localized. We can estimate the localized carrier density from the known gate efficiency of $2 \times 10^{11}\,\mathrm{cm}^{-2}\mathrm{V}^{-1}$ as calculated for the $n$-regime (Fig.~S5b). In the bulk $p$-regime, for $V_g^*=-2$\,V, the localized carrier density at the camelback is $4\times 10^{11}\,\mathrm{cm}^{-2}$. These localized carriers do not contribute to transport at the lowest temperatures ($20$\,mK).


\begin{figure}[hb!]
\includegraphics[width=0.9\linewidth]{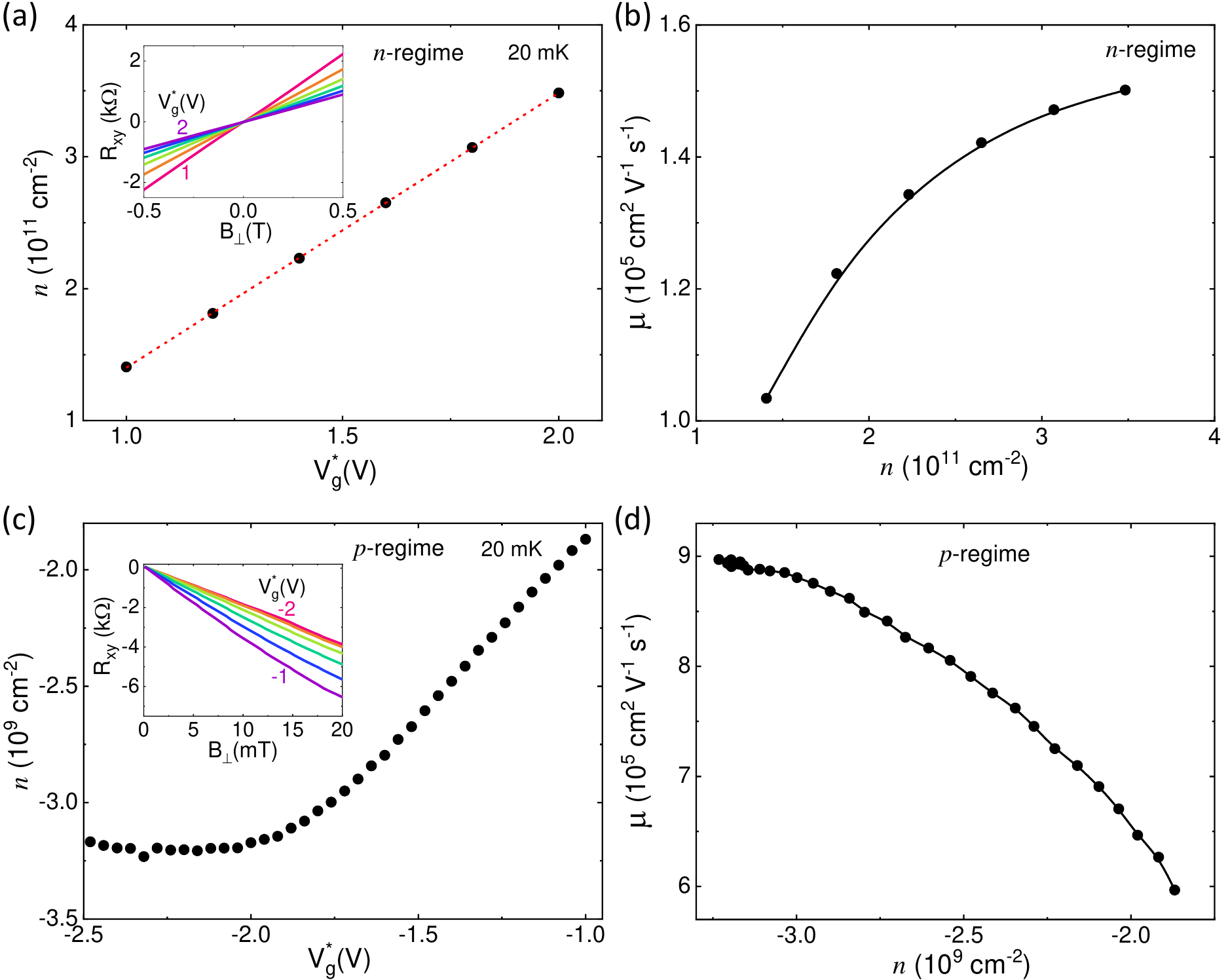}
\caption{\textbf{Density and mobility of charge carriers.}
(a) The carrier density $n$ as a function of normalized gate voltage $V_g^*$ at $20$\,mK in the bulk $n$-regime. The dashed red line is a linear fit to the data. The inset shows the transverse resistance $R_{xy}$ as a function of perpendicular magnetic field $B_\perp$ for $V_g^*$ ranging from $1$ to $2$\,V.
(b) The electron mobility $\mu$ as a function of $n$.
(c) The carrier density $n$ as a function of $V_g^*$ at $20$\,mK in the bulk $p$-regime. The inset shows $R_{xy}$ as a function of $B_\perp$ for $V_g^*$ ranging from $-1$ to $-2.5$\,V.
(d) The hole mobility $\mu$ as a function of $n$. The solid black curves are guides to the eye.}
\label{FigDensMob_Supp}
\end{figure}


\section{Magneto-resistance for Dev 2}
As shown in the band diagram of Fig.~1C in the main text, Dev 2 is an indirect-gap semiconductor with the energy at the camelback being higher than at $k=0$. We study the magneto-resistance of Dev 2 fabricated into a Hall bar of length $l=30$\,$\mathrm{\mu m}$ and width $w=10$\,$\mathrm{\mu m}$. The longitudinal resistance $R_{xx}$ shows a peak as function of gate voltage $V_g$, with maximum $R_{xx}$ being much greater than the expected $h/2e^2$ (in the QSH regime) due to the large dimensions of the device~\cite{KonigEA2007,BendiasEA2018}. For this device, we do not observe any early onset of the QH plateaus in the full gate voltage regime investigated here. The transverse resistance $R_{xy}$ shows the QH plateaus at perpendicular magnetic fields $B_\perp > 1$\,T in contrast to $30$\,mT for Dev 1.


\begin{figure}[hb!]
\includegraphics[width=0.9\linewidth]{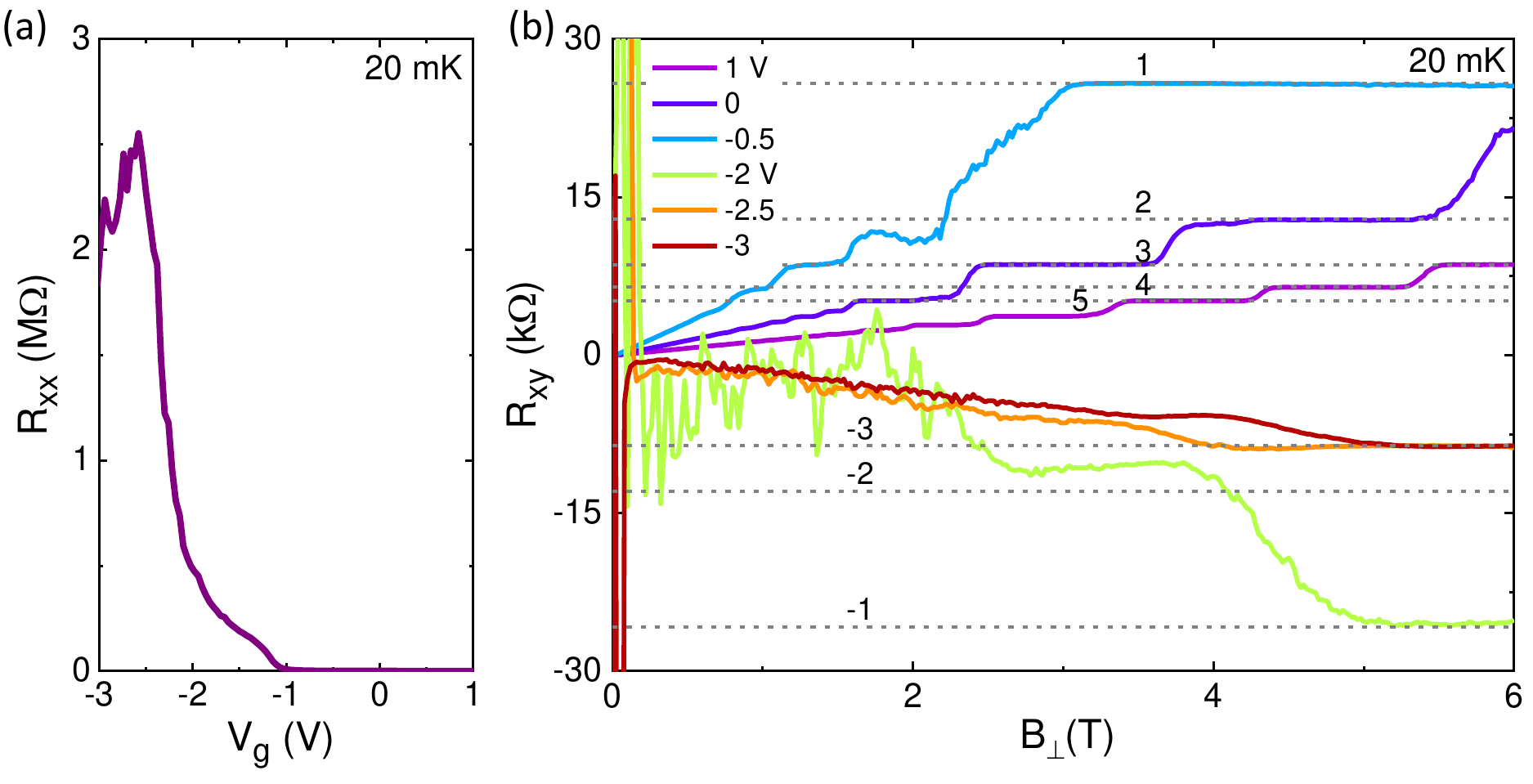}
\caption{\textbf{Magnetoresistance for Dev 2.}
(a) The longitudinal resistance $R_{xx}$ of Dev 2 as a function of gate voltage $V_g$ at $20$\,mK.
(b) The transverse resistance $R_{xy}$ of Dev 2 as a function of perpendicular magnetic field $B_\perp$ for different $V_g$ at $20$\,mK (see legend).}
\label{FigDev2_Supp}
\end{figure}


\section{$\kdotp$ calculations for a strip geometry}

\subsection{Modelling}

The band structures in this work are obtained from numerics involving the typical $\kdotp$ Hamiltonian for II-VI zincblende semiconductors \cite{Kane1957,NovikEA2005}.
We have used the standard $8\times8$ Kane Hamiltonian as described in Ref.~\cite{NovikEA2005}, involving the eight orbitals
\begin{equation}\label{eqnKaneBasis}
 \ket{\Gamma_6,+\tfrac{1}{2}},
 \ket{\Gamma_6,-\tfrac{1}{2}},
 \ket{\Gamma_8,+\tfrac{3}{2}},
 \ket{\Gamma_8,+\tfrac{1}{2}},
 \ket{\Gamma_8,-\tfrac{1}{2}},
 \ket{\Gamma_8,-\tfrac{3}{2}},
 \ket{\Gamma_7,+\tfrac{1}{2}},\text{\ and\ }
 \ket{\Gamma_7,-\tfrac{1}{2}}.
\end{equation}
We have used the Kane parameters for HgTe and CdTe from Ref.~\cite{NovikEA2005} as well. For Hg$_{1-x}$Cd$_{x}$Te, the Kane parameters are derived by cubic interpolation \cite{BerdingEA2000}.  For the Kane parameters for Hg$_{1-x}$Mn$_{x}$Te, we use the same values as for HgTe, with the exception of the conduction and valence band energies: We assume that the band gap decreases linearly as function of the concentration $x$, such that the band gap closes at $x=x_\mathrm{c}\approx0.064$ \cite{Furdyna1988}.
In addition to the standard Kane Hamiltonian, we include the appropriate terms for strain and the Zeeman effect \cite{NovikEA2005}. We do not consider explicitly the rather small effect of bulk inversion asymmetry (BIA), the strength of which is not known with sufficient precision for (Hg,Mn)Te, to the best of our knowledge.
However, we comment briefly on the influence of BIA terms in the last part of Supplementary Material. The exchange coupling describing the interactions between the carriers and the magnetic moments of the Mn is discussed in detail in Supplementary Section~\ref{SuppSecExchange}.

The Kane Hamiltonian is used to calculate the band structures in two geometrical configurations:
Firstly, for a \emph{bulk} geometry, we model the layer stack (see Methods, Material growth and characterization) by discretization in the growth direction $z$, making the Kane parameters and strain coefficients $z$-dependent.   In the remaining two directions (`in-plane'), the system has translational symmetry.  Diagonalization of this Hamiltonian thus yields energy eigenvalues $E_n(k_x,k_y)$ and eigenvectors $\ket{\psi_n(k_x,k_y)}$.  In view of clarity, we display dispersions $E_n(k)\equiv E_n(k/\sqrt{2},k/\sqrt{2})$ as a function of the single momentum variable $k$.
The choice of this parametrization is necessary in view of the lack of axial symmetry and the locations of the camelback maxima on the lines satisfying $k_x=\pm k_y$.
Secondly, for a \emph{strip} geometry, we discretize in one additional dimension ($y$).  Thus, there is only one remaining translational degree of freedom. The energy eigenvalues and eigenvectors are thus $E_n(k_x)$ and $\ket{\psi_n(k_x)}$, respectively.

For all our dispersions, we express energies relative to the zero energy $E\equiv0\meV$,  defined as the conduction band edge (four-fold degenerate $\Gamma_8$ orbital at $k=0$) of intrinsic (unstrained bulk) HgTe.  Generically, the edges of the bulk bands and the positions of the bulk gaps may vary depending on the material.  Thus, for the sake of comparing samples with various material compositions, we have decided to use a single reference energy rather than to centre all band gaps at $E=0\meV$.

Several properties of the eigenstates can be extracted from the eigenvectors $\ket{\psi_n(\vec{k})}$: Given an observable $O$, we calculate the eigenstate expectation values $\avg{O} = \bramidket{\psi_n(\vec{k})}{\hat{O}}{\psi_n(\vec{k})}$.  We represent these expectation values by colouring the dispersion curves. For example, for the strip geometry dispersions in Fig.~2C,D and 4C of the main text, the observable $O=y$ is used to determine whether the states are localized near the edges.  Also the orbital character ($O$ being a projection to the orbitals $\ket{\Gamma_6,\pm\frac{1}{2}}$, $\ket{\Gamma_8,\pm\frac{3}{2}}$, etc.)\ is a useful property that can be extracted from the eigenvectors.

The LL spectra (bulk geometry) are obtained by applying the Peierls substitution, $k_x\rightarrow k_x + e A_x / \hbar$, in the Landau gauge $A_x=-B_\perp y$. The calculation follows the standard procedure of replacing the canonical momentum operators by magnetic ladder operators followed by choosing an appropriate ansatz for the wave functions. The ansatz for the LL spinors in axial approximation is given by \cite{NovikEA2005}

{\small{}
\begin{align}
\vert \psi_{j}\left(z\right)\rangle=\begin{cases}
\left(\begin{array}{cccccccc}
f_{1}^{(j)}\vert j \rangle & f_{2}^{(j)}\vert j+1 \rangle & f_{3}^{(j)}\vert j-1 \rangle & f_{4}^{(j)}\vert j \rangle & f_{5}^{(j)}\vert j+1 \rangle & f_{6}^{(j)}\vert j+2 \rangle & f_{7}^{(j)}\vert j \rangle & f_{8}^{(j)}\vert j+1 \rangle\end{array}\right)^{T} & \text{for }j\geq1\\
\left(\begin{array}{cccccccc}
f_{1}^{\left(0\right)}\vert 0 \rangle & f_{2}^{\left(0\right)}\vert 1 \rangle & 0 & f_{4}^{\left(0\right)}\vert 0 \rangle & f_{5}^{\left(0\right)}\vert 1 \rangle & f_{6}^{\left(0\right)}\vert 2 \rangle & f_{7}^{\left(0\right)}\vert 0 \rangle & f_{8}^{\left(0\right)}\vert 1 \rangle\end{array}\right)^{T} & \text{for }j=0\\
\left(\begin{array}{cccccccc}
0 & f_{2}^{\left(-1\right)}\vert 0 \rangle & 0 & 0 & f_{5}^{\left(-1\right)}\vert 0 \rangle & f_{6}^{\left(-1\right)}\vert 1 \rangle & 0 & f_{8}^{\left(-1\right)}\vert 0 \rangle\end{array}\right)^{T} & \text{for }j=-1\\
\left(\begin{array}{cccccccc}
0 & 0 & 0 & 0 & 0 & f_{6}^{\left(-2\right)}\vert 0 \rangle & 0 & 0\end{array}\right)^T & \text{for }j=-2,
\end{cases}\label{eqA:llAnsatz}
\end{align}
}%
where $f_{i}^{(j)}\left(z\right)$ is the $z$-dependent envelope
function of the $j$-th LL with orbital component $i=1,\ldots,8$
enumerating the eight basis functions [Eq.~\eqref{eqnKaneBasis}]; $\vert j \rangle$ is the $j$-th eigenstate of a harmonic oscillator. In all bulk LL calculations, we apply the axial approximation which is needed to ensure that the LL index $j$ is a conserved quantity. This approximation reduces the numerical effort substantially, while the approximation error remains sufficiently small ($\approx1\,\mathrm{meV}$ at the camelback) \cite{NovikEA2005}.

\subsection{Numerical considerations}

The sizes $n\times n$ of the Hamiltonian matrices are given by $n=8n_z$ in the bulk geometry and by $n=8n_zn_y$ in the strip geometry, where $n_z$ and $n_y$ are the number of lattice points in the $z$ and $y$ direction, respectively.  For the discretization, we typically use a resolution of $0.25\nm$, where the discretization error is sufficiently small.  In the $z$ direction, the size of the well is $11\nm$ and we take into account $6\nm$ of each barrier (which is sufficient for the wave functions to decay to negligible values), so that $n_z\sim100$.  For the strip geometry, we restrict ourselves to a width $w=500$\,nm in order to keep the calculation tractable: There, we have $n_y\sim2000$, that leads to a matrix size of $n = 1.6\times10^6$.  We construct these large matrices in a sparse format and diagonalize them using a Lanczos algorithm
\cite{[{We have used the Python library SciPy, which provides an interface to the Lanczos methods implemented in ARPACK; see: }][{.}]ARPACK}
with the shift-and-invert method, that allows us to target a small number (typically $50$--$100$) of eigenvalues close to a target eigenvalue.

Although the strip width in the numerics ($w=500$\,nm) is significantly smaller than the actual sample size (typically $200\,\mathrm{\mu m}$), the numerical results remain suitable for studying the physics of the edge states.  The width is chosen to be larger than the localization length of these states, so that hybridization is negligible.  A small residual hybridization between edge states remains visible in the numerical results, but can be ignored for the experimental samples in view of their much larger size.

\clearpage

\section{Model for magnetic response of manganese}
\label{SuppSecExchange}

The \emph{exchange interaction} between the magnetic moments of the Mn ions and the carrier spins is typically understood in a mean field picture. The average Mn spin $\avg{\vec{m}}$ (proportional to the average magnetization) couples isotropically to the carrier spins
\begin{equation}\label{EqnHExchange}
  H_\mathrm{ex}
   = \sum_\alpha C^{(\alpha)}\avg{\vec{m}}\cdot \vec{\hat{S}}^{(\alpha)},
\end{equation}
where $\alpha$ labels the orbital sector (either $\Gamma_6$ or a combination of $\Gamma_8$ and $\Gamma_7$) and  $\vec{\hat{S}}^{(\alpha)}$ is the vector of spin operators in that sector. The coefficients $C^{(\alpha)}$ are phenomenological coupling constants that differ between the orbitals,
\begin{equation}
  C^{\Gamma_6} = -xN_0\alpha
  \quad\text{and}\quad
  C^{\Gamma_8,\Gamma_7} = -xN_0\beta,
\end{equation}
where $x$ is the Mn concentration (in Hg$_{1-x}$Mn$_{x}$Te), and $N_0\alpha=0.4$\,eV and $N_0\beta=-0.6$\,eV are empirically determined factors \cite{NovikEA2005}.

For the response of the average Mn spin to the external magnetic field, we assume the typical empirical law \cite{Furdyna1988,NovikEA2005,LiuEA2008PRL101}
\begin{equation}\label{EqnMagnetization}
 \avg{\vec{m}} = -S_0\frac{\vec{B}}{B_T} B_{5/2}\left(\frac{5}{2} \frac{g_{\mathrm{Mn}}\muB B_T}{\kB (T+T_0)}\right)
\end{equation}
where $B_T$ is the magnitude of the magnetic field and $\vec{B}/B_T$ its direction; furthermore, $B_{5/2}$ is the Brillouin function for spin-$\frac{5}{2}$, $g_\mathrm{Mn} = 2$ is the $g$ factor for Mn, $T+T_0$ is an effective temperature with an offset $T_0=2.6\,\mathrm{K}$, and $S_0=\frac{5}{2}$ is the effective total spin \cite{NovikEA2005,GuiEA2004}. We assume that the response is isotropic; the average magnetization is parallel to the magnetic field.

The characteristic magnetic field $B_\mathrm{ex}$ in \eqn\eqref{EqnHExchange}, defined by equating $\frac{5}{2} g_{\mathrm{Mn}}\muB B_T/\kB (T+T_0) = B_T/B_\mathrm{ex}$, has a lower limit of approximately $0.77\,\mathrm{T}$ at zero temperature.  In the field regimes discussed in this work, the argument $B_T/B_\mathrm{ex}$ of the Brillouin function is small, which justifies a linear approximation that yields
\begin{equation}\label{EqnMagnetizationLinear}
  \avg{\vec{m}}
  = -S_0\frac{\vec{B}}{B_T} B_{5/2}\left(\frac{B_T}{B_\mathrm{ex}}\right)
  \approx -\frac{7}{15}\frac{S_0}{B_\mathrm{ex}}\vec{B}
  = -\frac{7}{6}S_0\frac{g_{\mathrm{Mn}}\muB\vec{B}}{\kB (T+T_0)}.
\end{equation}
This argument demonstrates that we may assume that the average Mn spin $\avg{\vec{m}}$ is proportional to the magnetic field in good approximation.  In the context of the rotation experiments, where $\vec{B} = (B_x,0,B_z)$, the exchange interaction is thus proportional to
\begin{equation}\label{EqnHexSimple}
  H_\mathrm{ex}^{(\alpha)} \propto B_x \hat{S}^x + B_z \hat{S}^z
\end{equation}
within each orbital sector $\alpha$.

In order to demonstrate that the response of the states to the exchange interaction depends on the orbital content, we explore the following noteworthy scenarios:

%
\begin{itemize}

\item Subbands of heavy hole character, whose orbital character contains $\ket{\Gamma_8,\pm\frac{3}{2}}$ only, are unaffected by the spin operator $\hat{S}^x=\frac{1}{2}(\hat{S}^+ + \hat{S}^-)$,
\begin{equation}
 \bramidket{\Gamma_8,\pm\tfrac{3}{2}}{\hat{S}^x}{\Gamma_8,\pm\tfrac{3}{2}}
 = \bramidket{\Gamma_8,\pm\tfrac{3}{2}}{\hat{S}^x}{\Gamma_8,\mp\tfrac{3}{2}} = 0.
\end{equation}
Thus, the effective exchange Hamiltonian in this subspace is proportional to $B_z\hat{S}^z$ only. In other words, the eigenvalues of $H_\mathrm{ex}$ are proportional to $\pm\frac{1}{2}B_z$. There is no dependence on the in-plane component $B_x$.  This case applies to the lowest pair of conduction-band subbands near $k=0$ (subband character H1) as well as the other heavy-hole states which reside more deeply in the valence band (subband characters H2, H3, \ldots).

\item At $k=0$, the two E1 subbands are mixtures of electronic and light-hole orbitals, i.e., approximately $56\%$ of $\ket{\Gamma_6,\pm\frac{1}{2}}$ and $44\%$ of $\ket{\Gamma_8,\pm\frac{1}{2}}$.
In the $\Gamma_6$ sector, the eigenvalues of $H_\mathrm{ex}$ are isotropic, i.e., proportional to $B_T=\sqrt{B_x^2+B_z^2}$.  In the projection of the $\Gamma_8$ and $\Gamma_7$ sector to the subspace of $\ket{\Gamma_8,\pm\frac{1}{2}}$, we find an angular dependence of the eigenvalues,
\begin{equation}
  E_\mathrm{ex} \propto \pm\tfrac{1}{6}\sqrt{4 B_x^2+B_z^2}
  =\pm \tfrac{1}{6}B_\mathrm{T}\sqrt{1+3\sin^2 \theta}.
\end{equation}
The combination of the $\Gamma_6$ and $\Gamma_8$ sectors thus yields a pair of eigenvalues that is explicitly $\theta$-dependent, i.e., the exchange splitting is expected to be (weakly) angular dependent.

\item The states at the camelback are more complicated mixtures of the approximate form $\sqrt{0.7}\,\ket{\mathrm{E},\pm}+\sqrt{0.3}\,\ee^{\pm\ii\delta}\ket{\mathrm{O},\pm}$,
where $\delta$ is a mutual phase and
\begin{align}
  \ket{\mathrm{E},\pm} &= \phi_\mathrm{E}(z)\left(\sqrt{2/3}\,\ket{\Gamma_8,\pm\tfrac{1}{2}} \pm \ii\sqrt{1/3}\,\ket{\Gamma_8,\mp\tfrac{3}{2}}\right),\\
  \ket{\mathrm{O},\pm} &= \phi_\mathrm{O}(z)\ket{\Gamma_8,\pm\tfrac{3}{2}}\nonumber
\end{align}
are the `even' and `odd' components of these states, respectively.
Here, $\phi_\mathrm{E}(z)$ and $\phi_\mathrm{O}(z)$ denote even and odd wave functions in the spatial coordinate $z$, respectively.  The even and odd components are orthogonal, so they can be treated independently:  The odd sector is purely heavy hole, whose response to the exchange coupling we have discussed before.  In the even sector, the eigenvalues of $H_\mathrm{ex}$ are $E_\mathrm{ex}\propto\pm\frac{1}{18}\sqrt{40 B_x^2+B_z^2}$.  Combination of the two sectors also leads to an angular dependent exchange splitting.  The large coefficient for $B_x$ compared to that of $B_z$ suggests that this splitting strongly depends on the in-plane field component, which agrees qualitatively with the camelback splitting shown
in Fig.~4F 
of the main text.
\end{itemize}
%

For the latter two cases, we refrain from providing quantitative estimates because of the complicated dependence of the orbital content on other factors as well. The main conclusion is that the different responses to the exchange coupling can indeed be explained from the variation in orbital character.

\clearpage
\section{Further theoretical data for Dev 1 and Dev 2}
\label{SuppSecTheoryDev1Dev2}
\subsection{Landau level fan charts for Dev 1}
\label{SuppSecLLDev1}


\begin{figure}[b!]
\includegraphics[width=0.95\linewidth]{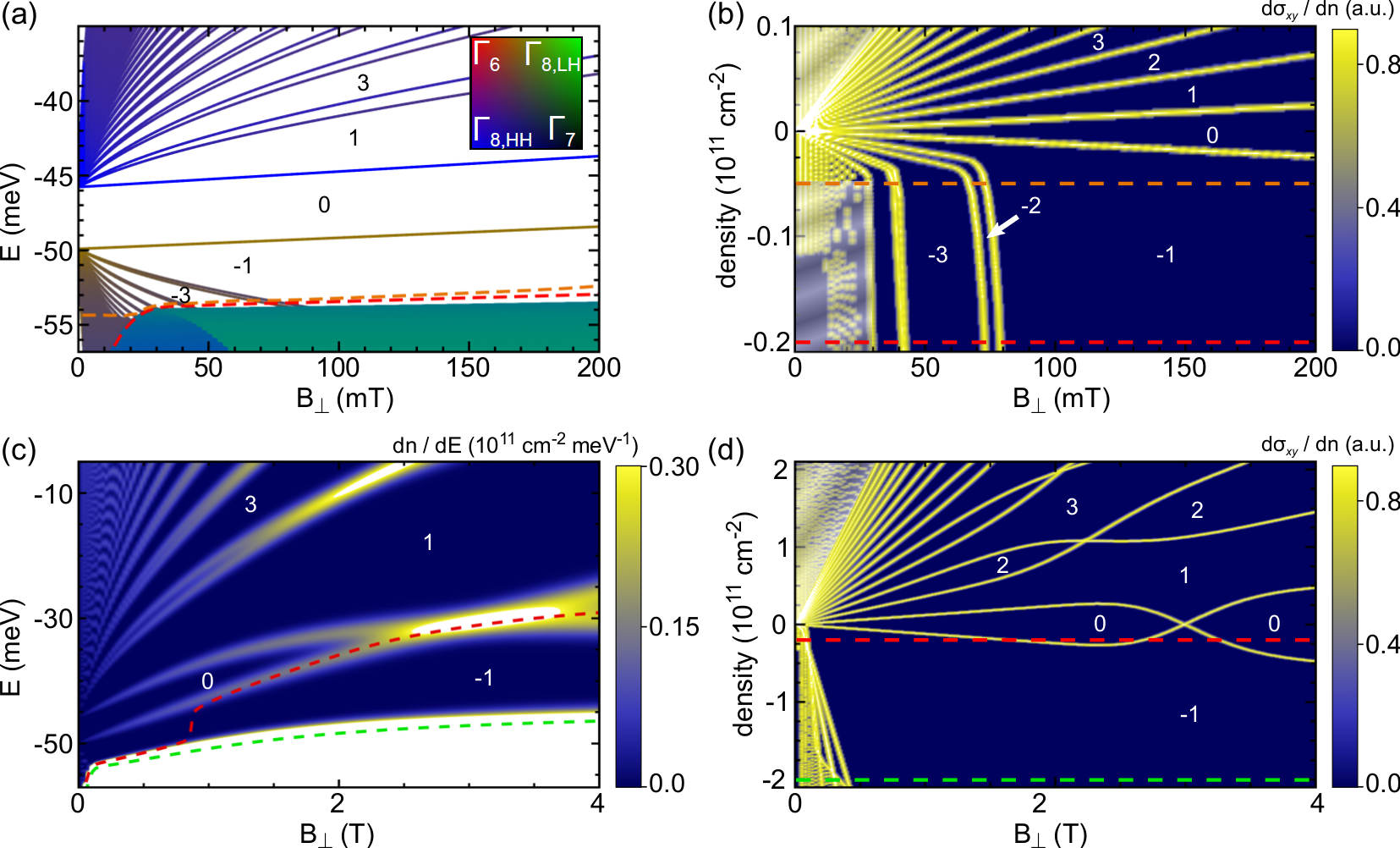}
\caption{\textbf{Band structure calculations for Dev 1.} (a) Bulk LL energies (no broadening) are shown as function of magnetic field; colour code indicates orbital character. (b) Different representation of the LL fan is presented, where the $y$-axis shows the carrier density and the colour code highlights $d \sigma_{xy}/dn$. In (a) and (b), dashed lines depict the evolution of the chemical potential as function of $B_\perp$ at constant density: $p = 0.05 \times 10^{11} \, \mathrm{cm}^{-2}$ (orange), and $p = 0.2 \times 10^{11} \, \mathrm{cm}^{-2}$ (red).
(c) The same LL fan chart for a much larger $B_\perp$-range is shown including broadening, where the colour indicates the DOS ($dn/dE$); a white colour implies that the DOS is out-of-scale.
(d) As in (b) but for the same magnetic field range as in (c). In (c) and (d), dashed lines depict the evolution of $\mu(B_\perp)$ at constant density: $p = 0.2 \times 10^{11} \, \mathrm{cm}^{-2}$ (red), and $p = 2 \times 10^{11} \, \mathrm{cm}^{-2}$ (green).
In (b) and (d), the shaded area at low magnetic fields marks a region below the computation limit. In all plots, numbers indicate the QH filling factors $\nu$. The label `$0$' indicates the quantum spin Hall state (edge states not shown).
}
\label{FigDev1_Supp_Theory}
\end{figure}


The LL spectrum for Dev 1 is shown in Fig.~S7a; the colour code indicates the orbital character. At $B_\perp = 0 \, \mathrm{T}$, the band structure is clearly inverted with the H1 subband (blue) being above the E1 subband (brown). While the conduction band looks quite conventional, the appearance of the valence band is dominated by a very dense collection of LLs. As described in the main text, we can attribute these LLs with a large LL index $j$ to the camelback (main text, Fig.~1B).

In order to calculate the chemical potential in the presence of a magnetic field $B_\perp$, we calculate the density of states $D(E)$ assuming that each LL is broadened with a Gaussian shape,
\begin{equation}
  D\left(E \right) = \frac{1}{2\pi l_{B_\perp}^2} \sum_{s} \, \frac{1}{\sqrt{2\pi \sigma^2}} \, \mathrm{exp} \left( - \frac{(E-E_s)^2}{2\sigma^2} \right),
\end{equation}
where the summation runs over all LLs $s$, $l_{B_\perp} = \sqrt{ \hbar / \mathrm{e} B_\perp}$ is the magnetic length, $E_s$ marks the energy of the respective LL and $\sigma$ determines the width of the LL broadening. Here, we take $\sigma=\sigma_0 \sqrt{B/B_0}$ with $\sigma_0 = 0.85 \, \mathrm{meV}$ and $B_0 = 1 \, \mathrm{T}$ \cite{NovikEA2005}. Due to the enormous amount of LLs arising from the camelback, the evolution of the chemical potential $\mu(B_\perp)$ for constant carrier densities is, for $p$-densities, determined by the shape of the camelback as indicated by the dashed lines in Fig.~S7a. QH transitions are observed whenever the chemical potential crosses a LL. The LL fan chart, shown in Fig.~S7a, is the basis for Fig.~3C of the main text, where, instead of the bulk LL spectrum, the density of states is highlighted.

Analogously to analyzing the evolution of $\mu(B_\perp)$ for various densities, one can also depict the LL fan chart in a $n(B_\perp)$ plot (Fig.~S7b), where $n$ is the total bulk carrier density and the colour code highlights $d\sigma_{xy}/dn$. Here, it is assumed that only extended states, located in the centre of each broadened LL,  can give rise to a change in $\sigma_{xy}$.  Since $n \sim V_g^*$, this form of depicting a LL fan chart is analogous to the experimental results presented in Fig.~3B of the main text. From Fig.~S7b, it is apparent that the camelback manifests itself as a large asymmetry between $n$- and $p$-densities.

To study also the high-field features of Dev 1, we show further results in Fig.~S7c,d. The characteristic LL crossing of an inverted band structure occurs for this specific sample at $B_\perp\approx 3 \, \mathrm{T}$. Due to pinning of the chemical potential to the camelback, we find that the $\nu=-1$ QH plateau can extend up to very large $B_\perp$ (Fig.~S7d). This is in good agreement with the experimental results  shown in Fig.~S3.

\subsection{Landau level fan charts and strip band structures for Dev 2}


\begin{figure}[bt]
\includegraphics{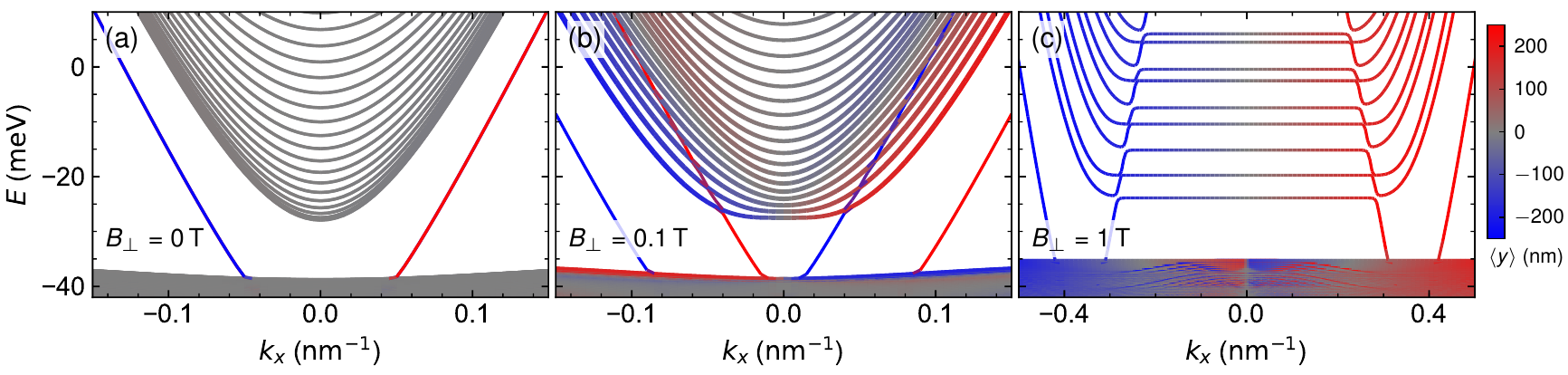}
\caption{\textbf{Band structure calculations on a strip for Dev 2.}
The band structures have been calculated on a strip of width $w=500$\,nm of the material of Dev 2,
(a) without magnetic field,
(b) with $B_\perp=0.1 \, \mathrm{T}$, and
(c) with $B_\perp=1 \, \mathrm{T}$.
We note the different scaling on the horizontal axis in (c).
}
\label{FigDev2_Ribbon}
\end{figure}



\begin{figure}[tbh!]
\includegraphics[width=0.9\linewidth]{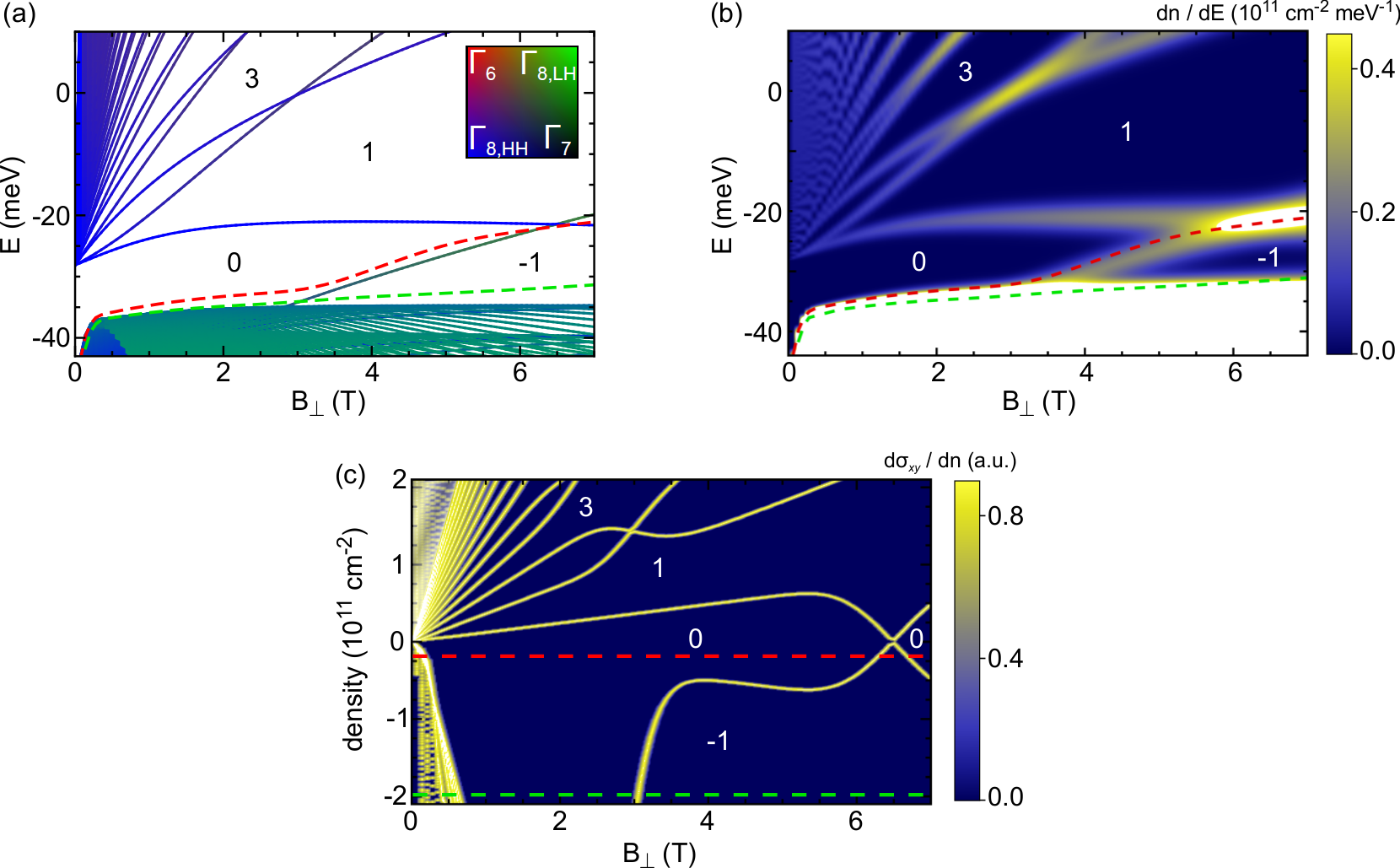}
\caption{\textbf{Band structure calculations for Dev 2.}
(a) Bulk LL energies (no broadening) are shown as function of magnetic field; colour code indicates orbital character.
(b) The same LL fan chart is shown including broadening, where the colour indicates the DOS ($dn/dE$); a white colour implies that the DOS is out-of-scale. The large white area arises due to the camelback.
(c) Different representation of the LL fan is depicted, where the y-axis shows the carrier density and the colour code highlights $d \sigma_{xy}/dn$. The shaded area at low magnetic fields marks a region below the computation limit. Numbers indicate the QH filling factors $\nu$, while the yellow lines depict transitions between adjacent LLs. In all plots, the dashed lines depict the evolution of the chemical potential as function of $B_\perp$ at constant density: $p = 0.2 \times 10^{11} \, \mathrm{cm}^{-2}$ (red), and $p = 2 \times 10^{11} \, \mathrm{cm}^{-2}$ (green).
}
\label{FigDev2_Supp_Theory}
\end{figure}


Further evidence in favour of this pinning mechanism can be gained by comparing theory and experiment for other device configurations.  For Dev 2, which is an indirect-gap semiconductor device unlike Dev 1, the Dirac point is buried deeply in the bulk valence band, as a strip calculation without magnetic field demonstrates (see Fig.~S8a). For small magnetic fields (e.g., $B_\perp=0.1 \, \mathrm{T}$, see Fig.~S8b), the LLs with small filling factors $\nu$ form close to the Dirac point, still deep in the bulk valence band.  Even a stronger field of $B_\perp=1 \, \mathrm{T}$ (Fig.~S8c) is insufficient to lift the $\nu=-1$ QH plateau above the camelback. These theoretical considerations corroborate the absence of the $\nu=-1$ QH plateau at small fields in the QSH regime in experiments (cf.\ Fig.~S6b).

To study also pinning to the camelback in the bulk $p$-regime, we provide Landau fans for Dev 2 in Fig.~S9.  Because of the indirect gap, the camelback penetrates into the QSH regime which is enclosed by the two characteristic LLs of an inverted band structure which cross at about $6.5 \, \mathrm{T}$.  Thus, LLs with small LL indices $j$ are covered completely by the camelback, preventing the early onset of the $\nu=-1$ plateau. The numerics (Fig.~S9a--c) demonstrate that  the chemical potential is instead pinned to the $\nu = 0$ plateau.  An onset to the $\nu=-1$ QH plateau should arise at $3$--$4 \, \mathrm{T}$, where the camelback separates from the QSH regime. We therefore do not show in detail the low-field regime ($B_\perp < 200 \, \mathrm{mT}$) for Dev 2.

\subsection{Influence of bulk inversion asymmetry}
Although our theoretical pinning model agrees qualitatively well with experiment for $p$-densities (Fig.~3B of the main text), there is a minor difference between theory and experiment on the quantitative level. In particular, the theory points to a broad $\nu=-3$ and a very narrow $\nu=-2$ QH plateau at low magnetic fields (Fig.~3C of the main text and Fig.~S7a), whereas in experiment $\nu=-2$ is the most visible QH plateau.
In the numerics, the even-odd difference arises since all valence band LLs with small $j$ come in almost-degenerate (small gap) pairs suggesting that only odd plateaus with $\nu=-1,-3,$ and $-5$ should be clearly resolved.  Each of these pairs is linked to two different LL spinors, i.e., $\psi_j$ and $\psi_{j+2}$, see Eq.~\eqref{eqA:llAnsatz}.


\begin{figure}[!tb]
\centering
\includegraphics[width=0.9\textwidth]{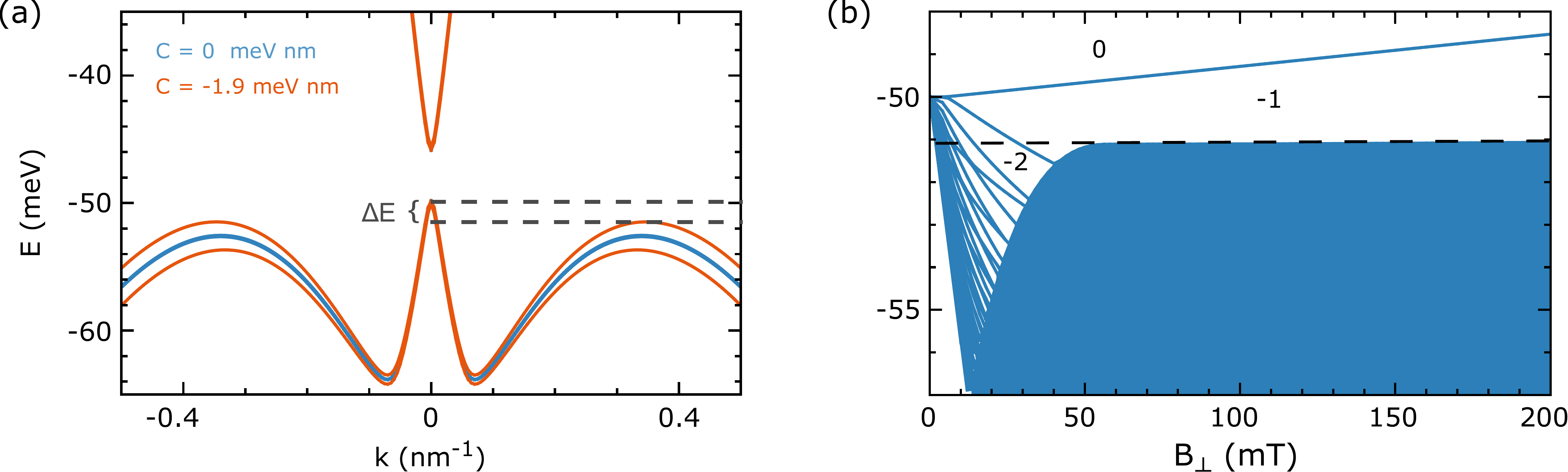}
\caption{\label{fig:BIA} \textbf{Influence of bulk inversion asymmetry for Dev~1.}
(a) Bulk inversion asymmetry (BIA) lifts the degeneracy already at $B_\perp=0$, causing a decrease of $\Delta E$ [cf. Fig.~S7(a)]. Blue and orange lines correspond to a linear BIA parameter 		$C=0$ and $-1.9 \meV \nm$, respectively (BIA Hamiltonian taken from Ref.~\cite{Winkler2003_book}).
(b) LL fan is shown including BIA, where numbers indicate QH filling factors $\nu$. The camelback crosses LLs with small LL indices at even smaller magnetic fields. Dashed line extrapolates camelback against $B_\perp=0$. Note that the $\nu=-2$ QH plateau is well resolved.
}
\end{figure}


This discrepancy may be resolved by considering a perturbation that couples these states and would split the LL pairs, thus leading to an increased width of all even plateaus ($\nu=-2,$ and $-4$).
Bulk inversion asymmetry, which has  been neglected so far for all theoretical figures, is a possible candidate: In linear approximation, the BIA Hamiltonian (see Ref.~\cite{Winkler2003_book}) couples states with $j$ to $j\pm2$. Thus, in presence of BIA, a linear combination of LL spinors, Eq.~\eqref{eqA:llAnsatz}, is an appropriate ansatz for the Kane Hamiltonian in magnetic fields.
\begin{align}
\vert \Psi^{\mathrm{E}} \left(z\right) \rangle	&=\sum_{j=-1}^{\infty}a_{2j}\vert \psi_{2j}\left(z\right)\rangle,\\
\vert \Psi^{\mathrm{O}} \left(z\right) \rangle	&=\sum_{j=-1}^{\infty}a_{2j+1}\vert \psi_{2j+1}\left(z\right)\rangle.
\end{align}
Here, $\Psi$ decomposes into an even and odd subspace labelled by $\mathrm{E}$ and $\mathrm{O}$, respectively, since only LLs with indices $j$ and $j+2$ hybridize.

Figure~S10(a) shows that including BIA in the band structure calculations lifts the degeneracy of the bulk valence bands already at $B_\perp=0$. This causes a decrease of the energetic difference $\Delta E$  between the valence band edge and the camelback maximum implying even smaller onset fields to QH plateaus in the bulk $p$-regime [cf. Fig.~S7(a) and~S10(b)].  The most important effect of including BIA is the broadening of all even plateaus, $\nu=-4,-2$. This shows that BIA is a potential candidate to explain the observed small discrepancy between theory and experiment. Here, we have used a realistic estimate for the strength of the BIA for (Hg,Mn)Te. However, since a precise value for the BIA in this material is unknown, an in-depth investigation into BIA is necessary in the future to obtain a better quantitative estimate.

\end{document}